\documentclass[journal]{IEEEtran}
\usepackage{graphicx}
\usepackage{amsmath}
\usepackage{multirow}
\usepackage{float}
\usepackage{bm}
\usepackage{amsfonts}
\usepackage{cite}

\ifCLASSINFOpdf
\else
\fi
\usepackage{cite}
\usepackage[numbers,sort&compress]{natbib}
\begin{document}
%

\title{NeuralDPS: Neural Deterministic Plus Stochastic Model with  Multiband Excitation for Noise-Controllable Waveform Generation}
%
%

\author{Tao~Wang,
        Ruibo~Fu,~\IEEEmembership{Member,~IEEE,}
       Jiangyan~Yi,~\IEEEmembership{Member,~IEEE,}
      Jianhua~Tao,~\IEEEmembership{Senior~Member,~IEEE,}
        and~Zhengqi~Wen
\thanks{T. Wang is with the National Laboratory of Pattern Recognition, Institute of
Automation, Chinese Academy of Science, Beijing 100190, China, and also
with the School of Artificial Intelligence, University of Chinese
Academy of Sciences, Beijing 100190, China (e-mail: tao.wang@nlpr.
ia.ac.cn).}
\thanks{R. Fu, J. Yi, J. Tao and Z. Wen are  with the National Laboratory of Pattern Recognition,
Institute of Automation, Chinese Academy of Sciences, Beijing 100190, China
(e-mail: \{ruibo.fu, jiangyan.yi, jhtao, zqwen\}@nlpr.ia.ac.cn).}

\thanks{Corresponding Author: Ruibo Fu, Jiangyan Yi,  Jianhua Tao.  E-mail: \{ruibo.fu, jiangyan.yi, jhtao\}@nlpr.ia.ac.cn.}}

%
%

\markboth{Journal of \LaTeX\ Class Files,~Vol.~14, No.~8, August~2015}%
{Shell \MakeLowercase{\textit{et al.}}: Bare Demo of IEEEtran.cls for IEEE Journals}
%



\maketitle

\begin{abstract}
The traditional vocoders have the advantages of high synthesis efficiency, strong interpretability, and speech editability, while the neural vocoders have the advantage of high synthesis quality. To combine the advantages of two vocoders, inspired by the traditional deterministic plus stochastic model, this paper proposes a novel neural vocoder named NeuralDPS which can retain high speech quality and acquire high synthesis efficiency and noise controllability. Firstly, this framework contains four modules: a deterministic source module, a stochastic source module, a neural V/UV decision module and a neural filter module. The input required by the vocoder is just the spectral parameter, which avoids the error caused by estimating additional parameters, such as F0. Secondly, to solve the problem that different frequency bands may have different proportions of deterministic components and stochastic components, a multiband excitation strategy is used to generate a more accurate excitation signal and reduce the neural filter's burden. Thirdly, a method to control noise components of speech is proposed. In this way, the signal-to-noise ratio (SNR) of speech can be adjusted easily. Objective and subjective experimental results show that our proposed NeuralDPS vocoder can obtain similar performance with the WaveNet and it generates waveforms at least 280 times faster than the WaveNet vocoder. It is also 28\% faster than WaveGAN's synthesis efficiency on a single CPU core. We have also verified through experiments that this method can effectively control the noise components in the predicted speech and adjust the SNR of speech.

\end{abstract}


%
\IEEEpeerreviewmaketitle

\section{Introduction}
%
%
%
%
\IEEEPARstart{A}{} vocoder   is used to generate a speech waveform with a parametric representation that can be converted back into a speech waveform \cite{zen2009statistical}.   Since it is convenient  to use the parameter representation to model speech, the vocoders are widely used in various tasks, such as statistical parametric speech synthesis \cite{king2011introduction}, voice conversion \cite{mohammadi2017overview}, singing voice synthesis \cite{saino2006hmm, nishimura2016singing} and speech coding \cite{flanagan1979speech, spanias1994speech}. A good vocoder is considered to restore the original waveform as much as possible through the parameter representation. However, since the parameter representation of speech is usually lossy, this brings great difficulties to model high-quality speech \cite{backstrom2017speech}.

\begin{figure}[tb]
    \centering 
    \includegraphics[width=7.5cm]{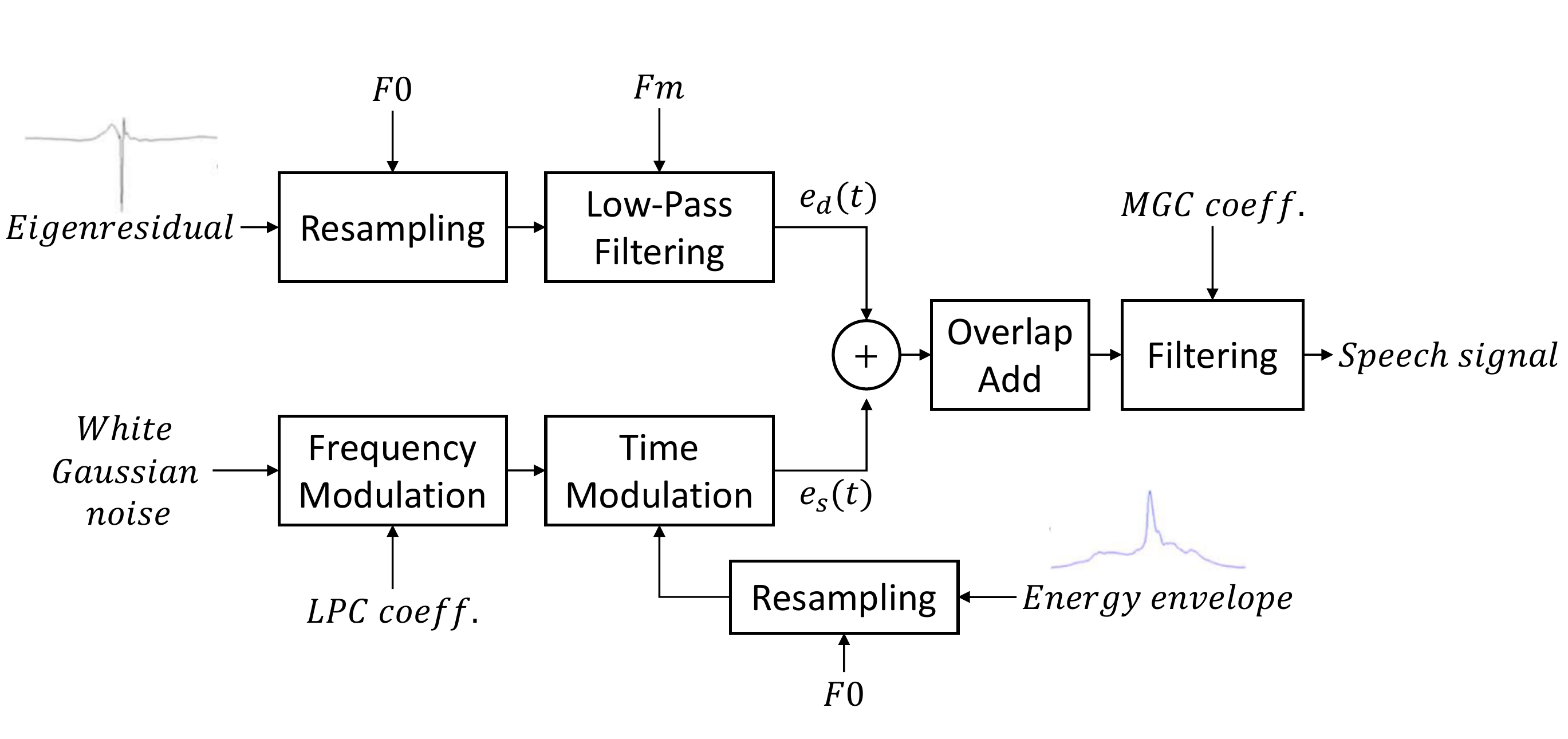}
    \caption{Workflow of the traditional deterministic plus stochastic vocoder. Input features are the target pitch F0 and the MGC filter coefficients. All other data is precomputed on a training dataset. The deterministic component $e_d(t)$ consists of the eigen residual resampled such that its length is twice the target pitch period. The stochastic part $e_s(t)$ is a white noise modulated by the autoregressive  model and multiplied in time by the energy envelope. The energy envelope is also resampled to the target pitch. Both components are then overlap-added.}
    \label{fig:dsp-1}
\vspace{-0.6cm}
\end{figure}

 In terms of statistical parameter speech synthesis (SPSS) \cite{king2011introduction}, the source-filter model \cite{vincent2007new} is used as the basic model of traditional vocoders. It assumes the speech is the convolution between the excitation signal and the filter. Among them, the excitation signal is the mixture of periodic  signal and aperiodic signal, and the filter models spectral envelope of speech which represents vocal tract  \cite{karjalainen2001generalized}. This model works well for speech.
The advantage of traditional vocoder is that it runs fast, and the output speech can be modified by modifying’’ the excitation signal and filter parameters \cite{morise2016world}.
But the disadvantage is  the low speech quality. There are two reasons for poor speech quality, one is the inaccurate estimation of acoustic features, and the other is the limitation of  the physical model \cite{airaksinen2018comparison,hu2013experimental}. Especially, the excitation signal is over-simple, which is usually simulated by the superposition of a periodic signal controlled by the fundamental frequency and a noise signal. Many researchers try to enrich the excitation signal to make the pronunciation more natural and accurate, such as the mixed excitation vocoder \cite{mccree1995mixed}, glottal vocoder \cite{hedelin1984glottal}, sinusoidal vocoder \cite{dolson1986phase,stylianou2001applying}, and deterministic plus stochastic model \cite{drugman2011deterministic}.  Since this paper is  inspired by the deterministic plus stochastic vocoder, we take it as an example. It is proposed to retain a more detailed excitation signal by substituting the pulse train with a residual signal which is shown in Fig. \ref{fig:dsp-1}. Quoting \cite{drugman2011deterministic}, the deterministic component can be defined as anything in the signal that is not noise (i.e., a perfectly predictable part), the stochastic component is defined as the residual signal obtained by subtracting the original speech signal's sinusoidal components. Both components are then overlap-added to obtain the total excitation signal. The final speech signal is generated by inputting  the total excitation signal to the Mel-log spectrum approximation filter. Although this vocoder has achieved competitive performance in traditional vocoders \cite{drugman2011deterministic}, the synthesized speech is still not natural enough because it is still too simple to use the traditional  signal processing method to simulate human pronunciation. However, its high generation efficiency and interpretable characteristics still have  efficient value.

With the development of machine learning \cite{lecun2015deep},  more breakthroughs have been reported for the vocoder. One of the pioneering works is WaveNet \cite{oord2016wavenet,shen2018natural}. In this work, the source-filter model is abandoned, and the speech generation is modeled by the convolutional neural network \cite{o2015introduction}.  On this basis, works such as WaveRNN \cite{kalchbrenner2018efficient} and SampleRNN \cite{mehri2016samplernn} are derived.
The generated speech of these vocoders can be comparable to human voices. Their common idea is to generate waveform points through autoregression form by only using a neural network.  Because of the autoregressive generation form, these vocoders can learn the temporal relationship between sequences well and synthesize high-quality speech. However, due to the larger number of speech waveform points, it is inefficient to generate speech in autoregressive form \cite{oord2018parallel}.
To improve the speed of synthesis, some parallel neural vocoders have been proposed. According to the principle, these vocoders can be classified into Flow-based model \cite{prenger2019waveglow}, GAN-based model \cite{yamamoto2020parallel,kumar2019melgan,su2020hifi}, VAE-based model \cite{peng2020non}, etc. \cite{oord2018parallel,kong2020diffwave}. Most of these vocoders have in common assumption that the speech can be mapped to a noise signal with a known distribution.  Based on this assumption, models with different principles have different optimization criteria. For example, WaveGAN \cite{prenger2019waveglow} optimizes the generative model through joint training an adversarial network \cite{yamamoto2020parallel}; WaveGLOW \cite{peng2020non} is trained using a single cost function: maximizing the likelihood of the training data.  Although these networks can produce high-quality speech, they often have low training efficiency or instability because the assumption that the speech is completely mapped from noise,  which will increase the burden of neural filter. Besides, since the neural network is hard to interpret, it’s impossible to edit the speech like the traditional vocoder.  

Some researchers try to combine traditional vocoder and neural network vocoders to achieve good synthesis efficiency and speech quality.  According to the source-filter model, speech generation is mainly divided into two modules: excitation module and filter module. The neural network vocoder mentioned earlier simulates the excitation source and the filter model at the same time. Some recent research works have found that one part adopts traditional methods, and the other part adopts a neural network, can generate high-quality speech while maintaining high-efficiency synthesis. For example, LPCNET \cite{valin2019lpcnet} uses a simple all-pole linear filter to represent the vocal tract response model. Then it uses a simple RNN network \cite{hochreiter1997long} to predict the residual signal, which can reduce the complexity of the neural network and improve the synthesis efficiency.
However, LPCNET is still in the form of autoregressive, which will face limitations when requires higher efficiency. To achieve parallel synthesis, another study attempts to improve from the excitation module.  The neural source filter module (NSF) \cite{wang2019neural} uses the fundamental frequency (F0) and the voiced and unvoiced (V/UV) flags to simulate the excitation source, which can bring prior knowledge to the neural vocoder. Similar work includes HiNet \cite{ai2020neural}, GlotNet \cite{juvela2019glotnet}, etc.
However, specific parameter estimates are required for the excitation model, such as F0 and V/UV flag.  On the one hand,  inaccurate F0 estimates or V/UV flag estimates often introduce very noticeable degradations in the synthesized speech. On the other hand, the traditional excitation source is still too simple, which may bring difficulties in reconstructing voice details.  Therefore, a better solution is to learn the excitation source automatically through  neural network, which can not only avoid the disadvantages of parameter estimation and oversimplification of the traditional source model, but also match the neural filter model more closely,  the effect of vocoders may be further improved.



This paper explores the approaches to improve the run-time efficiency of neural vocoder by modeling a more suitable waveform generation model through neural network. There are three main contributions.
Firstly, inspired by the traditional deterministic plus stochastic vocoder \cite{drugman2011deterministic},  which views the source model from a broader perspective as consisting of deterministic component and stochastic component, this paper proposes a baseline neural deterministic plus stochastic (NDPS-B) model. 
It has four parts: deterministic source module, stochastic source module, neural V/UV decision module, and neural filter module. These modules correspond to each module in the source-filter model, which has good interpretability and can synthesize natural speech which is similar to WaveNet.
Secondly, since the speech may have different proportions of deterministic signals and stochastic signals in different frequency bands, a multiband excitation strategy is adopted to enrich the excitation signal instead of employing a full band excitation signal. The NeuralDPS with multiband excitation (NDPS-MBE) can further improve the quality of synthesized speech.  Thirdly, a method to control the noise components of speech  based on  NeuralDPS is proposed. This method can control the signal-to-noise ratio (SNR) of speech by just controlling the  mask value predicted by the neural V/UV decision module, which can be applied in speech enhancement and other fields. 

The experimental results demonstrate that the proposed NDPS-MBE vocoder generates waveforms at least 280 times faster than the WaveNet vocoder and it is also 28\% faster than WaveGAN's synthesis efficiency on a single CPU core. The speech quality of NDPS-MBE is comparable to that of WaveNet.  In addition, we conduct detailed ablation experiments on NDPS-MBE to verify each module's role and improve the interpretability of the model. We have  also verified through experiments  that we can effectively control the noise components in the predicted speech and adjust  SNR of the speech.

This paper is structured as follows. 
The proposed vocoders (NDPS-B \& NDPS-MBE) and the way to control noise are described in Section \ref{sec:2}. After explaining the experiments in Section \ref{sec:3}, we draw a conclusion in Section \ref{sec:4}.
\vspace{-0.1cm}

\begin{figure*}[htp]
    \centering 
    \includegraphics[width=17cm]{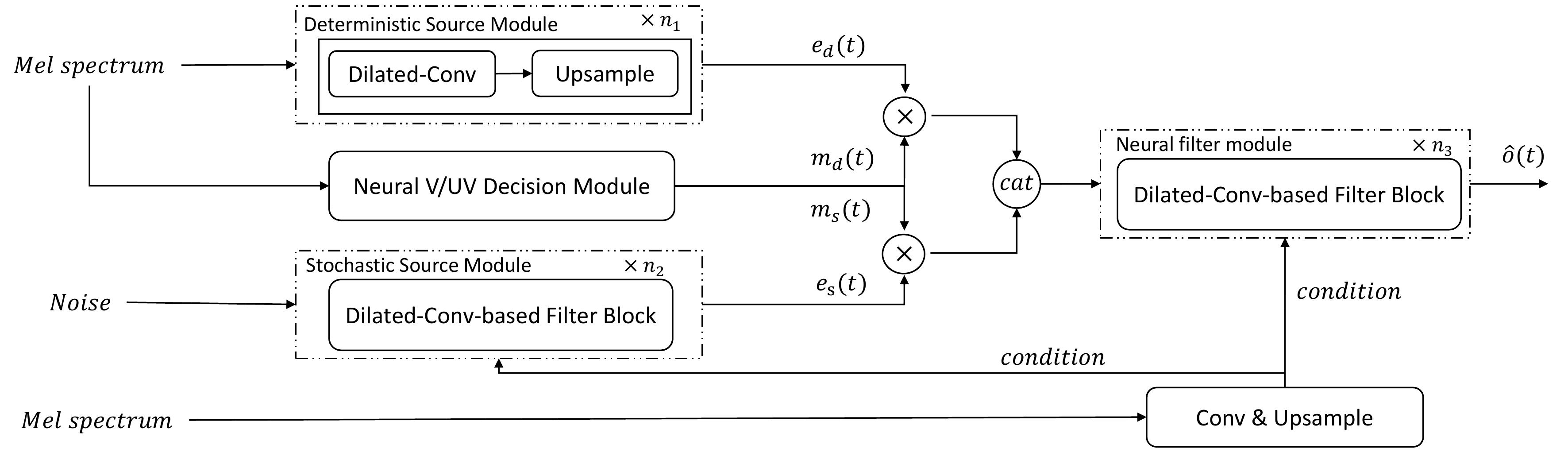}
    \caption{Structure of baseline NeuralDPS model (NDPS-B). Conv and Upsample denote convolution and upsamping layers, respectively.}
    \label{fig:figure1label3}
\vspace{-0.3cm}
\end{figure*}

\section{Neural deterministic plus stochastic model}\label{sec:2}
In this section, we first propose a  baseline neural deterministic plus stochastic (NDPS-B) model that is parallel in generation and straightforward in implementation. Based on the NDPS-B model, we add a multiband excitation strategy to enrich the excitation signal's information and further improve speech quality, denoted as DNPS-MBE.  Finally, based on the proposed model, we present a method to control the noise components in the synthesized speech.
\vspace{-0.3cm}

\subsection{Baseline NeuralDPS model (NDPS-B)} \label{sec:ndps-b}
 The baseline NeuralDPS (NDPS-B) shown in Fig. \ref{fig:figure1label3} is composed of four modules, which converts the acoustic feature sequence (for example, Mel spectrum \cite{zheng2001comparison}) $\textbf{C} = [C_1, \cdots, C_B]$ of length $B$ into a speech waveform $\widehat{\textbf{O}} = [\widehat{O}_1, \cdots, \widehat{O}_T]$ of length $T$. A deterministic source module is used to  generate the deterministic signal $\bm{e_d} = [{e_d}_1, \cdots, {e_d}_T]$ and a stochastic source module generates the stochastic signal $\bm{e_s} = [{e_s}_1, \cdots, {e_s}_T]$.  A neural V/UV decision module predicts the soft mask value of the deterministic signal and the stochastic signal. Finally, a neural filter module is designed to output the waveforms signal $\widehat{\textbf{O}}$  by combining the deterministic signal  $\bm{e_d}$ and the stochastic signal $\bm{e_s}$ with the mask information.

 \subsubsection{Deterministic Source Module}\label{sec:det}
 A deterministic signal is defined as anything that is not noise, in other words, anything that can be perfectly predicted from the past measured acoustic features. So the deterministic signal  $\bm{e_d} = [{e_d}_1, \cdots, {e_d}_T]$ is  
 generated from the acoustic features $\textbf{C}$.  Since the acoustic features are in the frequency domain, while the excitation signal is in the time domain, we use a series of upsampling and convolutional operations \cite{wang2019neural} to transform acoustic features $\bm{C}$ into excitation signals $\bm{e_d}$ in the time domain.
 A stack of residual blocks \cite{he2016identity} follows each transposed convolutional layer with dilated convolutions \cite{oord2016wavenet}. As the number of network layers increases, the receptive field of convolution layers will increase exponentially,  leading to better long-range correlation in the learned deterministic signal.

\subsubsection{Stochastic Source Module} \label{sec:sto}

In the traditional vocoders, the stochastic signal $\bm{e_s} = [{e_s}_1, \cdots, {e_s}_T]$  corresponds to a white gaussian noise $g(t)$ convolved with a filter model $h(t)$ whose time structure is controlled by an energy envelope $r(t)$ \cite{drugman2011deterministic}, which can be expressed as:
\begin{equation}
e_s(t) = r(t) \cdot [h(t) * g(t)]
\end{equation}
The use of $h(t)$ and $r(t)$ is required to account respectively for the spectral and temporal modulations.

In the stochastic source module, we use a multi-layer dilated convolutional network to model the filter $h(t)$ and use the acoustic features $\bm{C}$ as the condition information. Firstly, to combine the acoustic features information, we upsample the acoustic features through convolution and upsampling networks and get the hidden features $\bm{\widehat{C}} = [\widehat{C}_1, \cdots, \widehat{C}_T]$.
Then, we use the gaussian noise $\bm{g} = [g_1, \cdots, g_T]$ as input and expend it in dimension through an FF layer as $tanh(\bm{w}\cdot\bm{g} + \bm{b})$, where $\bm{w}$ is the transformation matrix and $\bm{b}$  is the bias. Thirdly, the expanded signal is processed by a dilated-conv layer, summed with the conditioned feature $\bm{\widehat{C}}$, processed by the gated activation unit based on $tanh$ and $sigmoid$, and transformed by two additional FF layers. This procedure is repeated $n$ times, and the dilation size of the dilated convolution layer in the $k$-th stage is set to $2 ^{k-1}$.
It should be noted that the number of repetitions $n$ does not need to be too large because stochastic signals are similar to noise and do not require too large receptive field.
The output after the $n$ stages is inputted to a FF layer with output dimension 1. The FF layer's output is the stochastic signal $\bm{e_s}$.

\subsubsection{neural V/UV decision module} \label{sec:uvu}
\begin{figure}[t]
    \centering 
    \includegraphics[width=7cm]{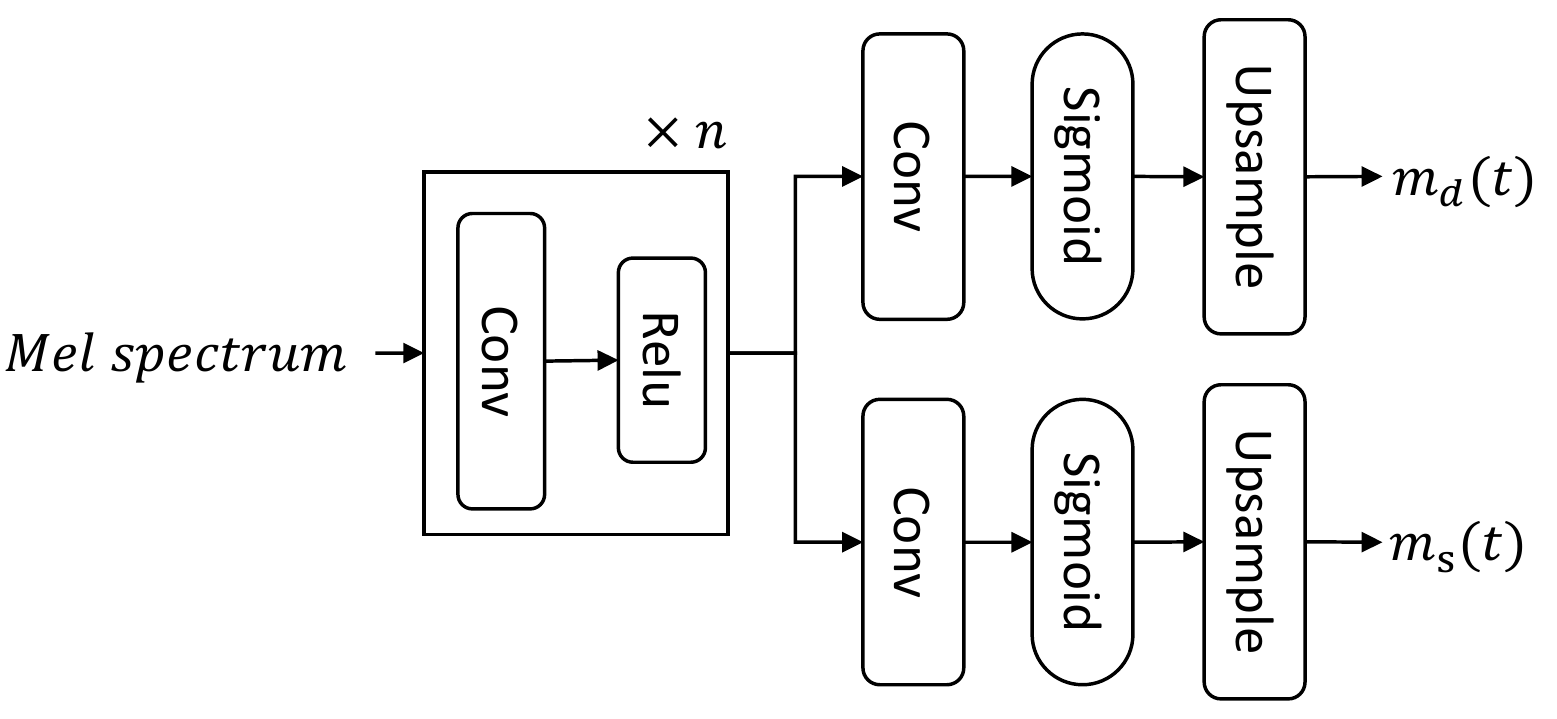}
    \caption{The structure of neural V/UV decision module.}
    \label{fig:uvu}
\vspace{-0.6cm}
\end{figure}
The neural V/UV decision module simulates the function of deciding the V/UV flag in the traditional vocoder. However, different from the  traditional V/UV decision module, this module predicts the soft mask values of deterministic signal $\bm{e_d}$ and stochastic signal $\bm{e_s}$. The range of the mask value is 0 to 1. At a specific moment, the greater the mask value of an excitation signal ($\bm{e_d}$ or $\bm{e_s}$), the greater the probability of selecting the excitation signal at that moment, which can help to obtain more accurate excitation signal. 

The structure of the neural V/UV decision module is shown in Fig. \ref{fig:uvu}. Firstly, we input the acoustic features $\bm{C}$ into several convolutional layers to get the context information $\bm{C^{*}}=[C^{*}_1, \cdots, C^{*}_B] $. Then, the context information $\bm{C^{*}}$ is input into two convolutional networks in parallel, where the output dimension of the convolutional network is 1, and the activation function is $sigmoid$.
The two sets of signals output by the two convolutional networks  are used as frame-level mask information, which is denoted as $\bm{m^f_d} =[{m^f_d}_1, \cdots, {m^f_d}_B] $ and $\bm{m^f_s} =[{m^f_s}_1, \cdots, {m^f_s}_B] $ with length $B$.
To map these frame-level mask value into the time dimension, we upsample the $\bm{m^f_d}$ and $\bm{m^f_s}$ and obtain the final mask value $\bm{m_d} =[{m_d}_1, \cdots, {m_d}_T]$ and $\bm{m_s} =[{m_s}_1, \cdots, {m_s}_T]$ with length $T$. The process can be expressed as:
\begin{equation}
\bm{m_d} = upsample(sigmoid(\bm{w_d}\cdot\bm{C^{*}} + \bm{b_d}))
\end{equation}
\begin{equation}
\bm{m_s} = upsample(sigmoid(\bm{w_s}\cdot \bm{C^{*}} + \bm{b_s}))
\end{equation}
where $\bm{w_d}$ and  $\bm{w_s}$ are the transformation matrix of the two convolutional networks, $\bm{b_d}$ and $\bm{b_s}$ are the bias.

We use the soft mask value $\bm{m_d}$ and $\bm{m_s}$ to mask the excitation $\bm{e_d}$ and $\bm{e_s}$. The masking mechanism can be expressed as $\bm{m_d} \odot \bm{e_d}$ and $\bm{m_s} \odot \bm{e_s}$, where $\odot$ denotes elements-wise multiplication. It should be noted that since the mask value is up-sampled from the mask information at the frame level, it is masked at the frame level. The advantage of this is to prevent the mask from being too fine-grained directly at every time instant.

\subsubsection{Neural Filter Module}
The function of neural filter module is to receive the masked excitation signal ($\bm{m_d} \odot \bm{e_d}$ and $\bm{m_s} \odot \bm{e_s}$)  and output the final waveform signal $\widehat{\textbf{O}}=[\widehat{O}_1, \cdots, \widehat{O}_T]$. The structure of neural filter module is the same as that of stochastic source module  introduced in the Section \ref{sec:sto}. 
Firstly, we concatenate the masked deterministic signal $\bm{m_d} \odot \bm{e_d}$ and the masked stochastic signal $\bm{m_s} \odot \bm{e_s}$ in the non-time dimension to obtain the total excitation signal, which is denoted as the $\bm{I} = [I_1, \cdots, I_T]$:
 \begin{equation}
   \bm{I}  = \text{cat} (\bm{m_d} \odot \bm{e_d} , \bm{m_s} \odot \bm{e_s} )
 \end{equation}
The concatenation rather than addition operation is used here, which can decouple the stochastic and deterministic signals intuitively. In this way, the input of the neural filter will be the independent stochastic signal and deterministic signal, rather than the total excitation signal after addition.

Then the total excitation signal $\bm{I}$ is input into the neural filter to obtain the final waveform signal $\bm{\widehat{O}}$, which can be expressed as:
\begin{equation}
  \bm{\widehat{O}} = NeuralFilter(\bm{I},\bm{\widehat{C}})
\end{equation}
 Where $NeuralFilter$  is the transformation process of the neural filter and $\bm{\widehat{C}}$ is the hidden features of acoustic features introduced in Section \ref{sec:sto}.

\subsubsection{Training Criterion} \label{lossfuc}
Two loss functions are defined between the predicted waveform $\bm{\widehat{O}}$ and the natural reference $\bm{O}$, including a multi-resolution amplitude spectrum loss and a waveform-domain adversarial loss. These  losses supervise the predicted waveform in  frequency domain and time domain  to ensure that the model will not overfit in any domain.

 The multi-resolution amplitude spectrum loss is based on multi-resolution STFT loss, which is the sum of the STFT losses with different analysis parameters (i e., FFT size, window size, and frameshift). For a single STFT loss, we minimize the spectral convergence $L_{sc}$ and log STFT magnitude $L _{mag}$ between the target   $\bm{O}$ and the reference $\bm{\widehat{O}}$.
\begin{equation}
L_{s c}(\bm{O}, \bm{\widehat{O}})=\frac{\||S T F T(\bm{O})|-\mid S T F T(\bm{\widehat{O}})\|_{F}}{\||S T F T(\bm{O})|\|_{F}}
\end{equation}
\begin{equation}
L_{\operatorname{mag}}(\bm{O}, \bm{\widehat{O}})=\frac{1}{N}\|\log |S T F T(\bm{O})|-\log \mid \operatorname{STFT}(\bm{\widehat{O}})\|_{1}
\end{equation}
where $\|\cdot\|_{F}$ and $\|\cdot\|_{1}$ represent the Frobenius and $L_{1}$ norms, respectively. $|S T F T(\cdot)|$ indicates the STFT function to compute magnitudes and $N$ is the number of elements in the magnitude. 
For the multi-resolution STFT objective function, there are $M$ single STFT losses with different analysis parameters. We average the $M$ operations through:
\begin{equation}
L_{s t f t}(G)=\mathbb{E}_{\bm{O}, \bm{\widehat{O}}}\left[\frac{1}{M} \sum_{m=1}^{M}\left(L_{s c}^{m}(\bm{O}, \bm{\widehat{O}})+L_{m a g}^{m}(\bm{O}, \bm{\widehat{O}})\right)\right]
\end{equation}
Where G represents the generator, that is, the vocoder we propose.  
We calculate the multi-resolution STFT loss between the output speech and the natural speech. In the back-propagation process, the loss can be propagated to the output of the model through STFT function, which is discussed in detail in paper \cite{wang2019neural}. Then the gradient at the model output can be propagated back to all the training parameters of the model to guide the parameter update \cite{lecun2015deep}.

To further improve the performance of proposed model, a discriminator is used to determine whether a waveform is true or false. Similar to WaveGAN \cite{oord2018parallel},
the waveform-domain adversarial loss is defined as:
\begin{equation}
L_{\mathrm{adv}}(G, D)=\mathbb{E}_{\boldsymbol{g} \sim N(0, I)}\left[(1-D(G(\bm{g},\bm{C})))^{2}\right]
\end{equation}
where $\bm{g}$ denotes the input gaussian noise. G represents the generator. D represents the discriminator. The input of D is speech waveform. The output of the discriminator is the speech type, that is, natural speech (represented by True) and speech synthesized by the generator (represented by False). 

Finally, the training criterion of the waveform generator is to minimize the combined loss function:
\begin{equation}
\mathcal{L}_{C o m b}=L_{s t f t}(G)+\lambda L_{\mathrm{adv}}(G, D) , (\lambda > 0)
\label{eq:totalloss}
\end{equation}
Where $\lambda$ is a hyperparameter.

\vspace{-0.1cm}
\subsection{NeuralDPS with multi-band excitation (NDPS-MBE)} \label{sec:ndps-mbe}

\begin{figure*}[htp]
    \centering 
    \includegraphics[width=17cm]{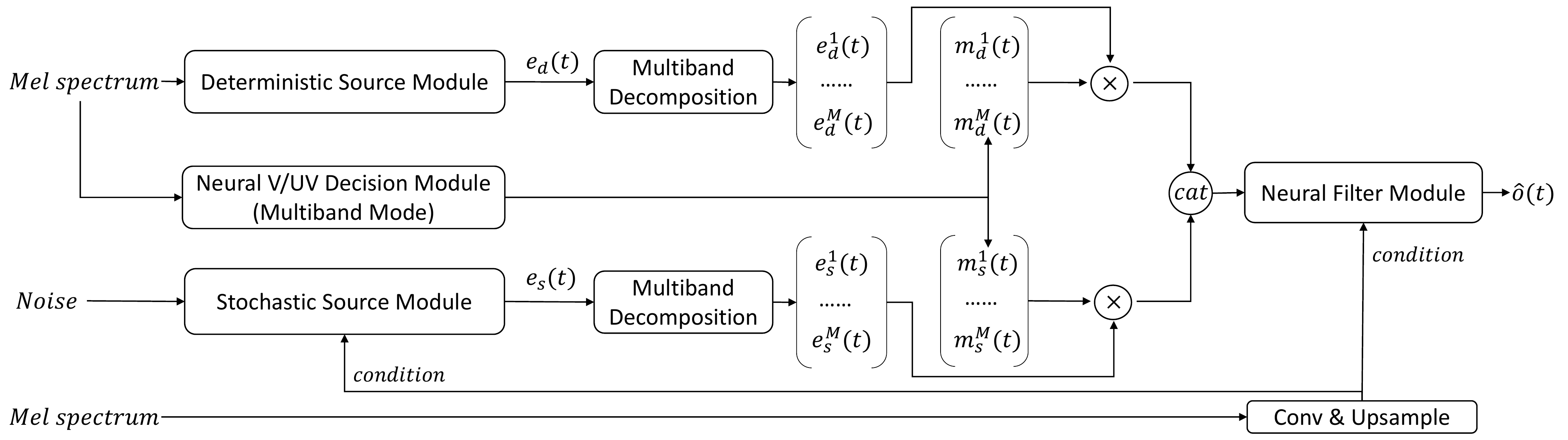}
    \caption{Structure of  NeuralDPS model with multiband excitation (NDPS-MBE). Conv and Upsample denote convolution and upsamping layers, respectively. The neural V/UV decision module in multiband mode outputs mask values of excitation signals ${e_d(t)}$ and $e_s(t)$ in different frequency bands.
}
    \label{fig:figure1label4}
\vspace{-0.2cm}
\end{figure*}

Although the neural V/UV decision module can learn adequate mask information similar to V/UV flag, the mask information is for the  full-band  speech signal. In the traditional vocoder, the disadvantage of employing  full-band V/UV flag to generate excitation signal is a "buzzy" quality especially noticeable in regions of speech which contain mixed voicing or in  regions of noise. This is because the full-band voiced/unvoiced excitation model produces excitation spectra consisting entirely of harmonics of the fundamental (voiced) or noise-like energy (unvoiced) \cite{griffin1988multiband}. Although the neural filter can alleviate the shortage of full-band excitation signal to a certain extent, it still brings additional burden to the neural filter module, which causes some details to be lost.

 A better strategy is to enhance the excitation signal by generating different frequency bands' excitation signals, an idea similar to the multiband excitation vocoder \cite{griffin1988multiband}.
Therefore, we divide deterministic signal $\bm{e_d}$ and stochastic signal $\bm{e_s}$ into several subbands according to different frequency bands by using the M-channel perfect-reconstruction (PR) cosine-modulated filter bank \cite{nguyen1994near}. It has emerged as an optimum filter bank with respect to implementation cost and design ease. To only decompose the excitation signal into multi subband signals, we just use the analysis filter of the M-channel PR cosine-modulated filter bank.

The impulse response of the analysis filters $h_k(n)$  is cosine-modulated versions of the prototype filter  $h(n)$, which can be expressed as:

\begin{equation}
\begin{aligned}
h_{k}(n)&=2 h(n) \cos \left((2 k+1) \frac{\pi}{2 M}\left(n-\frac{N-1}{2}\right)+(-1)^{k} \frac{\pi}{4}\right), \\  
&(0 \leq n \leq N-1  ,  0 \leq k \leq M-1)
\end{aligned}
\end{equation}

  \begin{table}[tp]
      \centering
          \caption{The parameters of prototype filter $h(n)$ when $M = 2$ or $M=4$.   $\beta$ can be determined by the desired stopband attenuation \cite{oppenheim1999discrete}, $N$ is the  filter order, and  $\omega_c$ is the cutoff frequency. }
  \begin{tabular}{c|ccc}
  \hline \hline & $\beta$ & $N$ & $\omega_c$ \\
  \hline M=2 & 9 & 62 & 0.142  \\
  \hline
   M=4 & 9 & 84 & 0.04\\
  \hline
  \end{tabular}
  \label{table:1}
\vspace{-0.3cm}
  \end{table}

 \begin{figure}[htbp]
      \centering 
      \includegraphics[width=7cm]{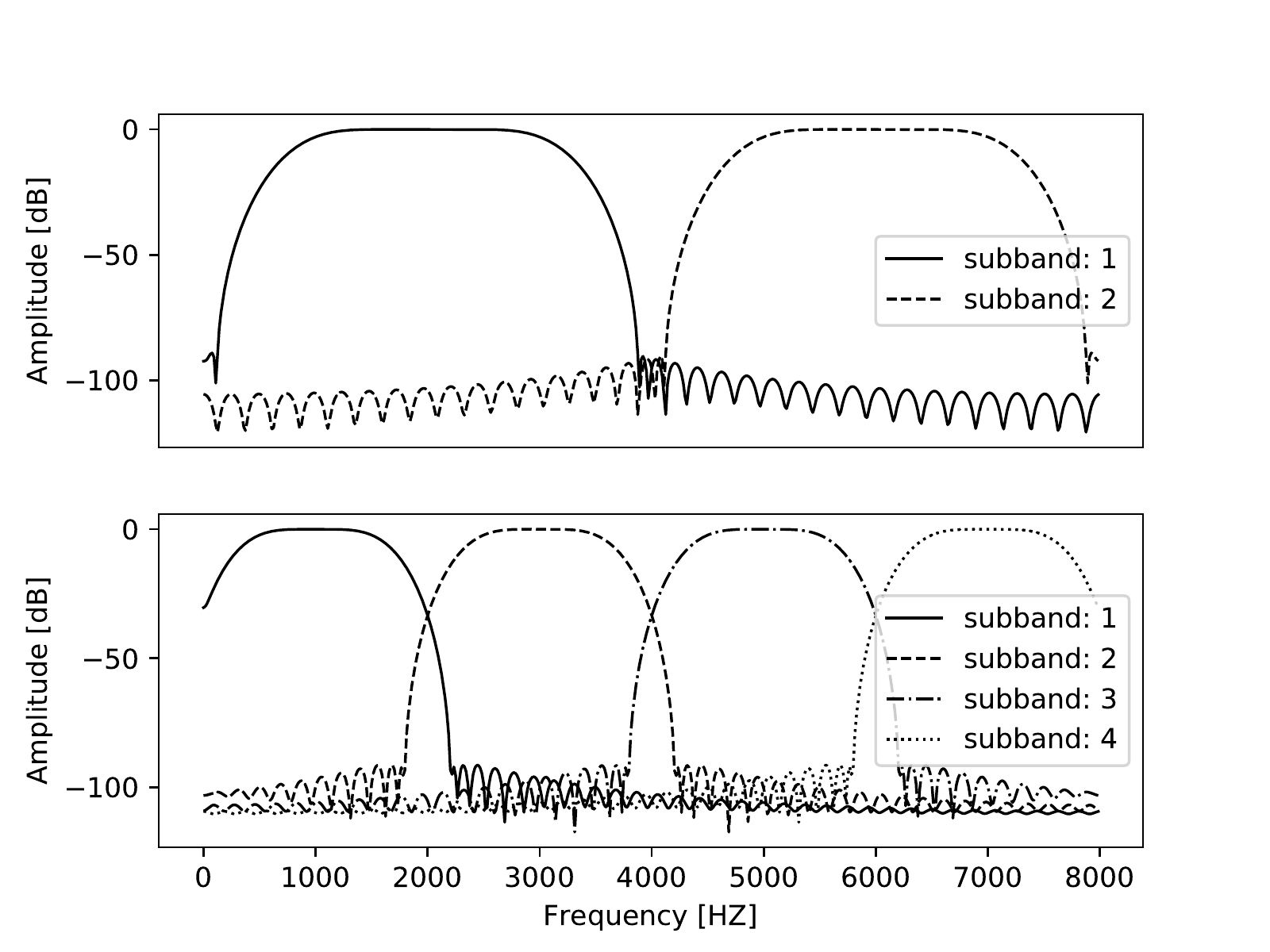}
      \caption{The impulse responses of the analysis filter $h_k(n)$ when $M = 2$ or $M=4$ , where $k$ is the order of subbands. }
      \label{fig:1}
\vspace{-0.1cm}
  \end{figure}

Where $h(n)$ is the Kaiser window function and the $N$ is the length of $h(n)$, $M$ is the number of subbands. 

The design of the prototype filter $h(n)$ is formulated as a problem of optimizing the cutoff frequency $\omega_c$, which is introduced in \cite{lin1998kaiser}. According to the paper, the coefficients of the prototype $h(n)$ can be expressed in closed form in terms of only three parameters,  $\beta$, filter order, $N$, and the cutoff frequency, $\omega_c$. The principle of our design is to make the overlap of different sub-bands as small as possible. When the number of subbands $M=2$ or $M=4$, the prototype filter parameters we designed are shown in Table \ref{table:1},  and the filter bank's impulse  response is shown in Fig. \ref{fig:1}.

Therefore, we can get different frequency band excitation signals by convolution of $h_{k}(n)$ $(k=1,2,\dots,M)$ and excitation signal $\bm{e_s}$ and $\bm{e_d}$. The process can be expressed as follows:
\begin{equation}
  \bm{e^{k}_d} = h_{k}(n) * \bm{e_d}, (k=1,2,\dots,M)
\end{equation}
\begin{equation}
  \bm{e^{k}_s} = h_{k}(n) * \bm{e_s}, (k=1,2,\dots,M)
\end{equation}
where the $\bm{e^{k}_d}$ and $\bm{e^{k}_s}$ are the final subband excitation signal in the $k$-th subband, respectively.

Correspondingly, we design the neural V/UV decision module into a multiband prediction mode. The whole framework  of NDPS-MBE is shown in Fig. \ref{fig:figure1label4}.
The  neural V/UV decision module in multiband mode does not output mask value of the full frequency band but predicts that of different subbands separately, which can be expressed as follows:
\begin{equation}
\bm{m_d ^{k}}= upsample(sigmoid(\bm{w_d ^{k}} * \bm{C^{*}} + \bm{b_d^{k}}))
\end{equation}
\begin{equation}
\bm{m_s ^{k}}=upsample(sigmoid(\bm{w_s ^{k}} *C^{*} + \bm{b_s^{k}}))
\end{equation}
where $k$ denotes the order of subbands, $\bm{w_d ^{k}}$ and  $\bm{w_s ^{k}}$ are the transformation matrix of the convolutional networks, $\bm{b_d^{k}}$ and $\bm{b_s^{k}}$ are the bias.

For the $k$-th subband, we use the $k$-th mask value $\bm{m_d^{k}}$ and $\bm{m_s^{k}}$ to mask the subband excitation signals  $\bm{e_d ^{k}}$ and $\bm{e_s ^{k}}$, and get the masked subband excitation signals $\bm{m_d^{k}} \odot \bm{e_d^k}$ and $\bm{m_s^{k}} \odot \bm{e_s^k}$. Finally, we concatenate all the masked subband excitation signal in the non-time dimension to obtain the total excitation signal $\bm{I}$, which can be expressed as:
\begin{equation}
  \bm{I} = cat(\bm{m_d^{k}} \odot \bm{e_d ^{k}},\bm{m_s^{k}} \odot  \bm{e_s ^{k}}), (k=1,2,\dots,M)
\vspace{-0.1cm}  
\end{equation}


\subsection{Editing speech by controlling noise}\label{sec:noisecontrol}
According to Section \ref{sec:ndps-b} and Section \ref{sec:ndps-mbe}, we have decoupled the vocoder as much as possible according to the function and modeled it with the neural network. Besides, we have also decoupled the deterministic signal and stochastic signal from the excitation signal. We can control the noise in speech by decoupling the excitation signal. For example, if there is too much noise in the original speech and we want to reduce the noise of the speech, we can reduce the mask value of the stochastic signal, so as to reduce the input of the stochastic signal. Finally, some noise in the speech will be removed.  Like the traditional vocoder, in which the speech can be edited according to the function of different modules, it is also possible for us to control the quality of speech based on our proposed neural vocoder.  This section presents two methods to control the SNR of speech by controlling the noise component in predicted speech. We take the NDPS-MBE vocoder as an example. The first method is to add a fixed constant $\mu_{noise} [-1,1]$ to all the predicted mask value $\bm{m_s^k}$   $(k=1,2,\dots,M)$ of stochastic signals. In this way,
the noise component of the predicted speech can be changed. The process can be expressed as:
\begin{equation}
  \bm{m_s^k} = \bm{m_s^k} + \mu_{noise}, -1 \leq \mu_{noise} \leq 1,
\end{equation}
\begin{equation}
  \bm{m_s^k} = clip(\bm{m_s^k},min=0,max=1)
\end{equation}
Where $k=1,2,\dots,M$. The function $clip$ is used to clip the value of $  \bm{m_s^k}$ between 0 and 1 to match the training process of the model.
Specifically, when $\mu_{noise} > 0$, more stochastic signal is introduced to the predicted speech, which will add more noise to the speech, thus reducing the signal-to-noise ratio of the predicted speech. On the contrary, when $\mu_{noise} < 0$, the predicted speech's noise will be reduced, which will improve the signal-to-noise ratio of speech.

Besides, since the excitation signals have been decomposed according to different frequency bands in model NDPS-MBE, another method is to  control specific subband's noise components. For example, we can reduce the low-frequency band's noise component to improve the speech quality because the low-frequency band's noise component significantly impacts the sound quality. Simultaneously, the high-frequency band's stochastic component is kept unchanged because the high-frequency part contains more details of speech.


\section{EXPERIMENTS AND RESULT ANALYSIS}\label{sec:3}
In the experiment, we adopt two speakers (the female speaker $slt$ and the male speaker $bdl$) in CMU-ARCTIC databases \cite{kominek2004cmu}, which contains English speakers with 16 kHz sampling rate. We chose 1000 and 66 utterances for each speaker to construct the training set and validation set, respectively. Another 66 utterances from the speakers not included in the training set and validation set are used as the test set.  

To evaluate the vocoder in SPSS-based systems,  since the CMU-ARCTIC databases is too small to train an end-to-end acoustic model, we also use the LJ Speech Dataset \cite{ljspeech17} to train Tacotron2  \cite{shen2018natural} models and vocoders. We chose 12600 and 250 utterances to construct the training set and validation set.  Another 250 utterances are used as the test set. The 80-dim Mel spectrogram used as acoustic features is extracted with Hann windowing, frameshift of 5 milliseconds, frame length of 50 milliseconds, and 1024-point Fourier transform.  Referring to the frame length setting in Tacotron, we set the frame length of the vocoder to  50 milliseconds. 

 Six vocoders are compared in our experiments. All models are trained and evaluated on a single  2080Ti GPU using pytorch \cite{paszke2017automatic} framework.  We choose the model from the following three perspectives. First, from the perspective of speech quality. we compared our model with the WaveNet model. Second, from synthetic efficiency and principle perspective. To improve the efficiency, some parallel neural vocoders have been proposed. We choose the model from the following three perspectives. First, from the perspective of speech quality. we compared our model with the WaveNet model. Second, from the perspective of synthetic efficiency and design principle. To improve the efficiency, some parallel neural vocoders have been proposed. We choose the vocoder, which is different from our model in some designs. For example, WaveGAN’s source signal is only a noise signal, while the source signal of our model is the learned excitation signal. Through the comparison of the two, the result can reflect the difference between whether there is a learned excitation signal. In addition, because the design of our model is mainly reflected in the source model, to reflect the difference between the source model of automatic learning and the source model of manual design, we take NSF as the comparison model.
The descriptions of these models are as follows \footnote{Examples of generated speech can be found at https://hairuo55.github.io/NeuralDPS.}.

\begin{itemize}
  \item \textbf{WORLD} The traditional vocoder which is based on the source-filter model. Because there is no open source about traditional deterministic plus stochastic model, and the effect of our implementation is not good, we use WORLD to represent  the traditional vocoder. The WORLD vocoder  is free software for high-quality speech analysis, manipulation and synthesis. It can estimate F0, aperiodicity and spectral envelope and also generate the speech like input speech with only estimated parameters. We use open source code  to implement it\footnote{https://github.com/mmorise/World.}.
  \item \textbf{WaveNet} A 16bit WaveNet-based neural vocoder which is trained using an open source implementation\footnote{https://github.com/r9y9/wavenet\_vocoder.}. To train 16kHz vocoder model, 3 upsampling layers with  upsampling rate \{5,4,4\} are adopted and other configurations remained the same as that of the open source implementation. The built model is a mixture density network, outputting the parameters for a mixture of 10 logistic distributions at each timestep and has 24 dilated convolutional layers which are divided into 4 convolution blocks. Each block contained 6 layers and their dilation coefficients are increased by an exponential of 2. The number of gate channels in gated activation units is 256. An Adam optimizer \cite{kingma2014adam} is used to update the parameters by minimizing the negative log likehood.
  \item  \textbf{WaveGAN} A 16bit Parallel WaveGAN neural vocoder which is trained using an open source implementation\footnote{https://github.com/kan-bayashi/ParallelWaveGAN.}. To train 16kHz vocoder model, 3 upsampling layers with  upsampling rate \{5,4,4\} are adopted and other configurations remained the same as that of the open source implementation. It has 30 dilated convolutional layers which are divided into 10 convolution blocks.  Each block contained 3 layers and their dilation coefficients are increased by an exponential of 2. The number of gate channels in gated activation units is 128. 
  \item \textbf{NSF} An NSF vocoder implemented by ourselves. The model structure is the same as that of the  WaveGAN.  The difference between the NSF and WaveGAN is the input excitation signal. The WaveGAN's input signal is noise signal, while the NSF's input signal is based on the  traditional source model.  Besides, the adversarial loss is added to the loss function to be consistent with the NeuralDPS in training criterion. We find it can improve the performance of the NSF model, too. The strategy of using separate source-filter pairs for the harmonic and noise components of waveforms is not adopted here.
  \item \textbf{NDPS-B} The proposed baseline NeuralDPS vocoder. To ensure the consistency of models' structure as much as possible, we set the parameters according to that of WaveNet and WaveGAN.  Since NeuralDPS can provide a more accurate source signal, we set the number of gate channels in the dilated-conv to 64, which is half smaller than WaveGAN. In the deterministic source module, the operation of dilated-conv and upsample is repeated 3 times ($n_1=3$), and the ratio of upsample is \{5,4,4\} in order.  In the stochastic source module, there is 3 dilated convolutional layers which are divided into 1 convolution block ($n_2=1$).  In the neural filter module, there is 24 dilated convolutional layers which are divided into 8 convolution blocks ($n_3=8$). Each block of the stochastic source module and the neural filter module contains 3 layers and their dilation coefficients are increased by an exponential of 2.
 The number of residual channels is 64 and the number of skip channels is 64, which is same as that of WaveGAN.
   For the loss function, three set of STFT configurations (FL,FS,FN), i.e., (320,80,512), (640,160,1024) and (1960,256,2048), are used for the amplitude spectrum loss.  The discriminator for generation adversarial learning
consists of ten layers of non-causal dilated 1-D convolutions with leaky ReLU activation function ($\alpha=0.2$). The strides are set to one and linearly increasing dilations are applied for the 1-D convolutions starting from one to eight except for the first and  last layers. The discriminator is consistent with that of the WaveGAN. An Adam optimizer is used to update the parameters. The init learning rate is 0.001 and the learning rate decreased exponentially. 
  \item \textbf{NDPS-MBE} The proposed NeuralDPS with multiband excitation.  The difference between NDPS-B and NDPS-MBE is that NDPS-MBE has a process of decomposing the excitation signal according to different frequency bands. We decompose deterministic signal $\bm{e_d}$ and stochastic signal $\bm{e_s}$ into 2 subbands ($M=2$) respectively based on NDPS-B.  Therefore, the neural V/UV decision module has four convolution layers at the end of the model to output mask value of the four subbands. The input dimension of neural filter is changed from 2 to 4. When the subbands' number $M=2$, the configuration of the prototype filter $h_n(t)$ is shown in the Table \ref{table:1}. Other configurations are consistent with model NDPS-B. 
\vspace{-0.1cm}
\end{itemize}

\begin{table*}[t]
    \centering
        \caption{OBJECTIVE EVALUATION RESULTS OF WORLD, WAVENET, WAVEGAN, NDPS-B AND
NDPS-MBE ON THE TEST SETS OF TWO SPEAKERS WHEN USING NATURAL ACOUSTIC FEATURES AS INPUT}
\begin{tabular}{cc|ccccc}
\hline \hline & & WORLD & WaveNet & WaveGAN  & NDPS-B & NDPS-MBE \\
\hline & SNR(dB) & 0.5569 & \textbf{5.4132}  & 5.1660  & 4.7130 & 4.7384 \\
& LAS-RMSE(dB) & 0.9037 & \textbf{0.7534} & 0.7665 &0.7864 &0.8005 \\
slt &  MCD(dB) & 3.7937 & 1.2479 &  \textbf{1.1860} & 1.2866 & 1.3073 \\ 
& F0-RMSE(cent) & 11.0624&9.6556&10.9926&9.8536& \textbf{9.3965} \\
& V/UV error(\%) & 4.1273&3.3154&3.5954&3.2924& \textbf{2.8610} \\
\hline
 & SNR(dB) &-0.3357& \textbf{2.8234}&2.0199&1.7275&1.9913
\\
& LAS-RMSE(dB) & 0.9456&0.7659& \textbf{0.7507}&0.8333&0.8172\\
bdl &  MCD(dB) &2.4200&1.3950&\textbf{1.2638}&1.3651&1.3863\\
& F0-RMSE(cent) & 21.6607&16.6038&17.5691&14.7835&\textbf{14.4492}\\
& V/UV error(\%) &  12.1593&7.5321&8.0851&8.1384&\textbf{7.2799} \\

\hline \hline
\end{tabular}
\label{table:2}
\vspace{-0.1cm}
\end{table*}

\begin{figure*}[htp]
    \centering 
    \includegraphics[width=16cm]{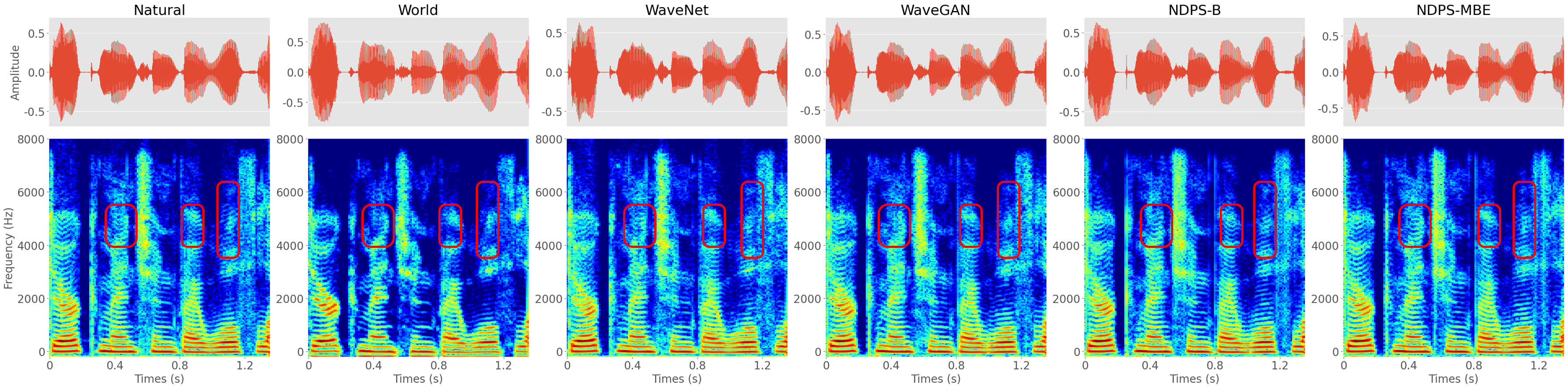}
    \caption{The waveforms and spectrograms of natural speech and the speech generated by different vocoders when using natural acoustic features as input for an example sentence.}
    \label{fig:figure1label}
\vspace{-0.4cm}
\end{figure*}

\subsection{Comparison between NDPS and Some Existing Vocoders}\label{sec:ndps-com}

This section compares the performance of our proposed NDPS-B and NDPS-MBE vocoder with three existing representative vocoders, including WORLD, WaveNet and WaveGAN by objective and subjective evaluations.

First, we compare the distortions between natural speech and the speech synthesized by these vocoders when using natural acoustic features as input. Five objective metrics in \cite{tamamori2017speaker} are adopted here, including signal-to-noise ratio ($\text{SNR}$), which reflects the distortion of waveforms, root MSE (RMSE) of LAS (denoted by $\text{LAS-RMSE}$), which reflected the distortion in the frequency domain, Mel-cepstra distortion ($\text{MCD}$), which described the distortion of Mel-cepstra, MSE of F0 which reflected the distortion of F0 (denoted by $\text{F0-RMSE}$), and $\text{V/UV error}$ which is the ratio between the number of frames with mismatched V/UV flags and the total number of frames. Specifically, to calculate the gap between synthesized speech and natural speech at the waveform point level, we use SNR metric for evaluation. According to the description in paper \cite{tamamori2017speaker}, we first calculate the SNR value for each frame of speech. In this process, the natural speech is regarded as signal, and the gap between synthesized speech and natural speech is regarded as noise. Time shift between $\pm200$ points that maximizes the correlation between natural speech and synthesized speech is calculated for each frame, and applied it to the windowed synthesized speech. The final SNR is calculated for each frame and averaged over total frames.

The results on the test sets are listed in Table \ref{table:2}. Firstly, the World vocoder has the worst performance on all metric for both speakers, which shows the advantages of neural network modeling.
Secondly, on the  $\text{SNR}$, WaveNet achieved the best performance in the two data sets, 
while in the frequency domain metrics $\text{LAS-RMSE}$  and $\text{MCD}$,  WaveNet and WaveGAN achieve better performance. This phenomenon shows that our method cannot achieve the best in both time domain loss and frequency domain loss.
Thirdly, by comparing the $\text{F0-RMSE}$ and $\text{V/UV error}$ metrics,  the NDPS-MBE achieves the best performance among the five vocoders which implies the advantage of using a neural V/UV decision module, which can help to learn more accurate F0 information and V/UV flag.
Finally, by comparing model NDPS-B and model NDPS-MBE, it can be found that the model NDPS-MBE is better than model NDPS-B on some metric, such as the $\text{SNR},\text{F0-RMSE}$ and $\text{V/UV}$ metric. But in terms of frequency domain metrics  $\text{LAS-RMSE}$  and $\text{MCD}$, model NDPS-B is better. We speculate that this is because the full band excitation  can provide complete information, so as to help the model optimize to lower spectrum loss. The multiband excitation signal can help the model improve the SNR in the time domain and learn more accurate fundamental frequency information.  It is worth mentioning that, the SNR values in the two test sets in Table II are at different levels. We attribute this to the difference in the amount of noise contained in the two test sets. Specifically, the less the noise component of the data set, the smaller the gap between the synthesized speech and the natural speech at the waveform point level, and the higher the SNR value. A similar phenomenon appears in paper \cite{tamamori2017speaker,ai2020neural}, which can also prove that it is caused by the difference between data sets and has nothing to do with the calculation process.


Fig. \ref{fig:figure1label} shows the waveforms and spectrograms of natural  speech and the speech generated by different vocoders when using natural acoustic features as input. It can be found that the neural vocoder restores the overall waveform contours better than the traditional vocoder world. In addition, by comparing the spectrum of speech synthesized by different vocoders, it can be found that the WaveNet and WaveGAN models will blur in the high frequency part, while the speech synthesized by NDPS-B or NDPS-MBE model has a more clear spectrum in the high frequency part. This proves the advantage of combining the physical background of traditional vocoder with neural network.

\begin{table}[t]
    \centering
        \caption{REAL TIME FACTORS (RTFS) ON GPU AND CPU AND NUMBER OF MODEL
PARAMETERS OF COMPARED VOCODERS}
\scalebox{0.9}{
\begin{tabular}{c|cccc}
\hline \hline Vocoder & WaveNet & WaveGAN & NDPS-B & NDPS-MBE \\
\hline RTF (GPU) & 170.217 & 0.015 & \textbf{0.010} & 0.011  \\
RTF (CPU) & 603.892 & 3.006 & \textbf{2.142} & 2.180  \\
Params(M) & 43.748 & 17.099 & \textbf{11.832} & 11.836 \\
\hline \hline
\end{tabular}}
\label{table:rft}
\vspace{-0.2cm}
\end{table}

To evaluate the run-time efficiency of different neural vocoders, real-time factor (RTF), defined as the ratio between the time consumed to generate speech and the duration of the generated speech, is utilized as the measurement. The smaller the RTF, the higher the efficiency. In our implementation, the RTF value is calculated as the ratio between the time consumed to generate all test sentences using a single GeForce RTX 2080Ti GPU or a single CPU core and the total duration of the test set. The results are listed in Table \ref{table:rft}. Our models significantly improve generation efficiency compared to the autoregressive model WaveNet. Specifically, the RTF of model NDPS-B on CPU is 2.142, which means that the NDPS-B needs 2.142s operation time to generate one-second speech on CPU. Compared with the RTF value (3.006)  of WaveGAN on CPU, the efficiency of our model is improved by 28\%.
 In addition, we compare the size of the model parameters. The last column of Table \ref{table:rft} counts the parameters of each model. It can be observed that the total parameters of the NDPS-B and NDPS-MBE are much smaller than WaveNet and they are 30\% smaller than WaveGAN's parameters. This is because the learned excitation signal reduces the burden of the neural filter, thereby greatly reducing the amount of parameters of the neural filter.

\begin{figure}[htp]
    \centering 
    \includegraphics[width=7cm]{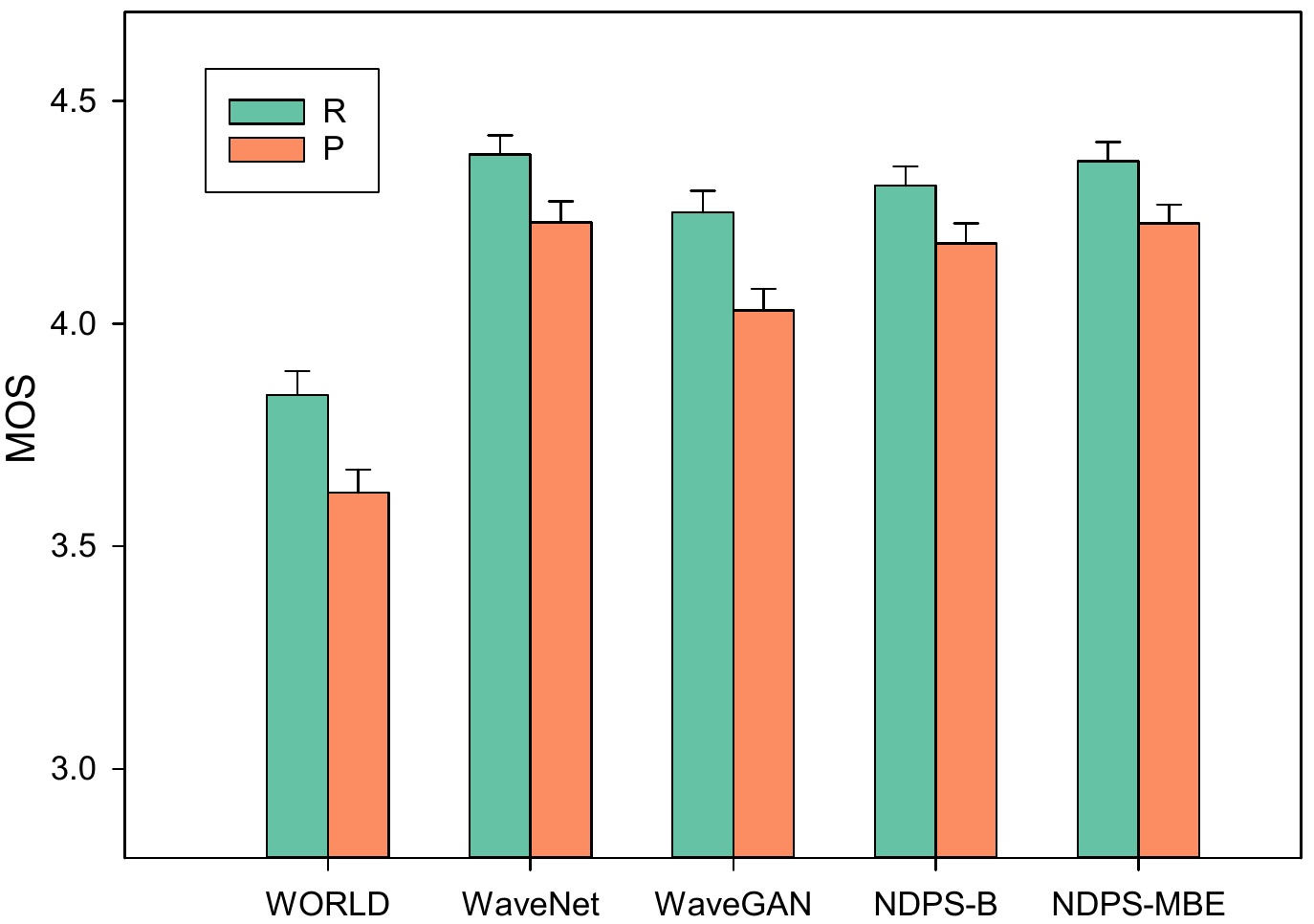}
    \caption{The MOS score with 95\% confidence intervals of the five vocoders for speaker $slt$. "R" stands for using natural acoustic features as input and "P" stands for using predicted acoustic features as input. The predicted acoustic features are predicted by Tacotron2.}
    \label{fig:mos}
\end{figure}

We conduct the Mean Opinion Score (MOS) listening test for speech quality on the test set regarding the subjective evaluation. We keep the text content consistent among different models to exclude other interference factors and examine speech quality. In addition, to evaluate the performance of the vocoder in the TTS task, we input text into Tacotron2 \cite{shen2018natural} to predict 80-dim Mel acoustic features. The predicted acoustic features are input into the vocoder to obtain the final speech, represented by “P” in Fig. \ref{fig:mos}. The structure of Tacotron2 is the same as that described in paper \cite{shen2018natural}. We convert text into phonemes as input to Tacotron2.
Twenty listeners participated in the evaluation. In each experimental group, 20 parallel sentences are selected randomly from the test sets of each system. Fig. \ref{fig:mos} shows the MOS score of each system.  The results show that the model NDSP-B and the model NDPS-MBE are better than the WaveGAN model. This is because the WaveGAN only uses noise as excitation signal, while model NDPS-B and NDPS-MBE can learn more rich excitation signal from noise and acoustic features.
 By comparing the WaveNet vocoder and NDPS-MBE vocoder, it can be observed that there is a small gap between the two vocoders in subjective evaluation.  Although our proposed NDPS-MBE vocoder achieved similar performance with WaveNet, its run-time efficiency is about 280 times higher on a single CPU core, as shown in Table \ref{table:rft}. Besides, by comparing model NDPS-B and model NDPS-MBE, it can be found that the multi-band excitation strategy can further improve the speech quality. By using multiband excitation, the neural filter can flexibly select the excitation signal according to different frequency bands signal rather than only selecting the full band excitation signal.
\vspace{-0.2cm}
\subsection{Comparison between NDPS and NSF}

In this section, we compare the performance of proposed NDPS-B and NDPS-MBE with the NSF vocoder.
The main difference between our model and NSF model is adopting different excitation signals. The NSF model artificially designs the excitation signal through the $\text{F0}$ and the V/UV flag. The NDPS-B and NDPS-MBE's excitation signal are learned from neural network. By comparing with the NSF model, we can find out which excitation signal is more suitable for the neural vocoder. 
We add an adversarial loss for joint training and find that it can improve the NSF model's performance.

\begin{table}[t]
    \centering
        \caption{OBJECTIVE EVALUATION RESULTS OF  NDPS-B, NDPS-MBE AND
NSF ON THE TEST SETS OF SPEAKERS $SLT$ WHEN USING NATURAL
ACOUSTIC FEATURES AS INPUT}
\scalebox{1.05}{
\begin{tabular}{c|ccc}
\hline \hline    & NDPS-B & NSF & NDPS-MBE  \\
\hline  SNR(dB)  &4.7130& 2.4063   &\textbf{ 4.7384} \\
 LAS-RMSE(dB)    & 0.7864  & \textbf{0.7047}  & 0.8005\\
  MCD(dB)    & \textbf{1.2866 } &  1.3332        & 1.3073 \\ 
 F0-RMSE(cent)   & 9.8536 & 11.7124      & \textbf{9.3965}   \\
 V/UV error(\%)  & 3.2924  &  2.8985     & \textbf{2.8610 }\\
\hline \hline
\end{tabular}}
\label{table:nsf}
\vspace{-0.2cm}
\end{table}

The objective results on the $slt$ speaker dataset are listed in Table \ref{table:nsf}. It shows model NDPS-MBE is better than NSF on SNR metric, which means that the learned excitation signal can help the vocoder to learn more natural speech in the time domain. While the model NSF is better than model NDPS-MBE in frequency domain metric  $\text{LAS-RMSE}$. This shows that  the NSF model could get better performance in frequency domain metric because of the artificial designed excitation signal, which reduces the complexity of the model. However, in the synthesized speech of NSF vocoder, there are often too many harmonics in the high frequency part of the voiced segment, which affects the speech quality. This may be due to the simple source model of NSF, resulting in overfitting of the model. 
The NDPS-B and NDPS-MBE models are better than NSF in terms of $\text{F0}$ and $\text{V/UV}$ metrics.  This shows that compared with the manually designed excitation signal, the automatically learned excitation signal is helpful to learn more stable and accurate F0 information.

\begin{table}[t]
\centering
\caption{AVERAGE PREFERENCE SCORES (\%) ON SPEECH QUALITY BETWEEN
NDPS-MBE  AND NSF MODEL WHEN USING
NATURAL ACOUSTIC FEATURES AS INPUT, WHERE N/P STANDS FOR “NO
PREFERENCE” 
}
\begin{tabular}{cccc}
\hline
\hline
NDPS-MBE & NSF & N/P & p \\ \hline
\textbf{39.75} & 32.25   & 28.00 &  0.027\\ \hline \hline
\end{tabular}
\label{table:nsfabx}
\vspace{-0.1cm}
\end{table}

In addition to objective metrics, we conduct ABX tests on our proposed model and NSF model. In each subjective test, twenty sentences randomly selected from the test set are synthesized by two comparative vocoders. At least 20 listeners evaluate each pair of generated speech. The listeners are asked to judge which utterance in each pair has better speech quality or no preference. To calculating the average preference scores, the $p$-value of the $t$-test is used to measure the significance of the difference between two vocoders. The results are listed in Table \ref{table:nsfabx}. Obviously, the performance of NDPS-MBE is better than that of NSF ($p$\ \textless0.05). This may be attributed to the poor listening feeling caused by excessive high-frequency harmonics of the waveforms generated by NSF.  This also shows that the excitation signal automatically learned by the neural network is better than the excitation signal designed by signal processing method.

\subsection{Ablation Test on NDPS-MBE}
In this section, we conduct ablation experiments on NDPS-MBE vocoder to explore the role of each module.

\subsubsection{The adversarial loss}
As introduced in Section \ref{lossfuc}, a combination of amplitude spectrum loss and an adversarial loss is used to train the NDPS-MBE vocoder. The amplitude spectrum loss can help the model converge quickly and learn adequate semantic information. In this section, we will explore the effect of the adversarial loss.  Based on the NDPS-MBE model, we remove the adversarial loss from the total loss function $\mathcal{L}_{C o m b}$ in Eq. \ref{eq:totalloss} to train the vocoder which is denoted as \textbf{NDPS-MBE-woGAN}.

\begin{table}[t]
    \centering
        \caption{OBJECTIVE EVALUATION RESULTS OF  NDPS-MBE AND
NDPS-MBE-woGAN  ON THE TEST SETS OF SPEAKERS $SLT$ WHEN USING NATURAL
ACOUSTIC FEATURES AS INPUT}
\begin{tabular}{c|cc}
\hline \hline    & NDPS-MBE & NDPS-MBE-woGAN  \\
\hline  SNR(dB)  & \textbf{4.7384} & 3.0206 \\
 LAS-RMSE(dB)    &  0.8005  & \textbf{0.7655}   \\
  MCD(dB)    & 1.3073 &  \textbf{1.1906 }       \\ 
 F0-RMSE(cent)   & 9.3965 & \textbf{9.2356}    \\
 V/UV error(\%)  & 2.8610  &  \textbf{2.8504}    \\
\hline \hline
\end{tabular}
\label{table:gan}
\vspace{-0.2cm}
\end{table}

The objective metrics on the $slt$ speaker dataset are shown in the Table \ref{table:gan}. It can be observed that removing the adversarial loss led to the degradation of $\text{SNR}$ metric. However, some metrics in the frequency domain ($\text{LAS-RMSE}$, $\text{MCD}$) are better. This is because only optimizing the spectrum  will make the model lack phase information. However, the adversarial loss is to judge whether the input speech of the discriminator is natural speech (represented by True) or speech generated by the generator (represented by False). The input of the discriminator is in the time domain and contains phase information. Thus, the adversarial loss makes up for the disadvantage that the spectrum loss does not contain phase information.

\begin{table}[t]
\centering
\caption{AVERAGE PREFERENCE SCORES (\%) ON SPEECH QUALITY BETWEEN
NDPS-MBE  AND NDPS-MBE-woGAN MODEL WHEN USING
NATURAL ACOUSTIC FEATURES AS INPUT, WHERE N/P STANDS FOR “NO
PREFERENCE”
}
\begin{tabular}{cccc}
\hline
\hline
NDPS-MBE & NDPS-MBE-woGAN & N/P & p \\ \hline
\textbf{80.50} &  9.25  & 10.25 &  \textless0.01 \\ \hline \hline
\end{tabular}
\label{table:ganabx}
\end{table}

We also conduct a subjective ABX test to compare NDPS-MBE with NDPS-MBE-woGAN. The result is listed in Table \ref{table:ganabx}.  It can be observed that the quality of  speech is obviously ($p\ \textless0.01$) decreased after removing the adversarial loss.

In summary,  the adversarial loss helps the model to learn the phase information of the speech, which makes up for the defect of only optimizing the amplitude spectrum loss, and makes the synthesized speech more natural.

 \begin{table*}[t]
 \centering
 \caption{AVERAGE PREFERENCE SCORES (\%) ON SPEECH QUALITY BETWEEN
NDPS-MBE-woS,  NDPS-MBE-woD and  NDPS-MBE WHEN USING
NATURAL ACOUSTIC FEATURES AS INPUT, WHERE N/P STANDS FOR “NO
PREFERENCE” 
}
 \begin{tabular}{c|ccccc}
 \hline
 \hline
  & \textbf{NDPS-MBE-woS} & \textbf{NDPS-MBE-woD} & \textbf{NDPS-MBE} & \textbf{N/P} & \textbf{p} \\ \hline
 NDPS-MBE-woS vs NDPS-MBE  & 14.25  & --  & 77.25 & 8.50 &   \textless  0.01 \\
NDPS-MBE-woD vs  NDPS-MBE  &  --   & 22.75 & 36.50  & 40.75 & \textless  0.01 \\
 NDPS-MBE-woS vs  NDPS-MBE-woD   & 18.50    & 65.75   & -- & 15.75 &  \textless  0.01 \\ \hline \hline
 \end{tabular}
 \label{table:wods}
\vspace{-0.3cm}
 \end{table*}

\begin{figure}[htp]
    \centering 
    \includegraphics[width=8.2cm]{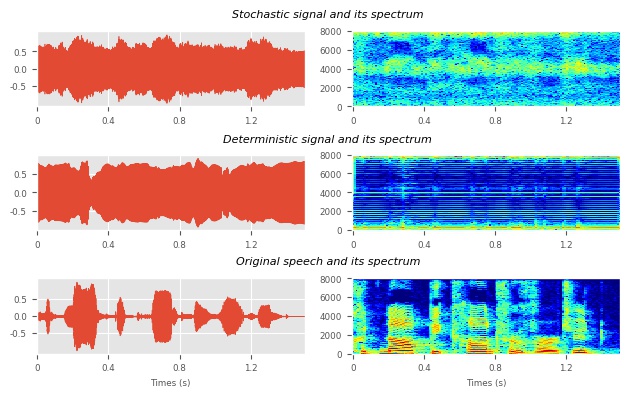}
    \caption{The waveforms and spectrograms of  the learned deterministic signal, stochastic signal and their original speech. The deterministic signal has obvious periodicity, while the stochastic signal does not have periodicity. Since the  deterministic signal and stochastic signal have different scales, we normalize their time domain waveform signals to [0,1]}
    \label{fig:det_sto}
\end{figure}
\subsubsection{The deterministic and stochastic source modules}
As introduced in Section \ref{sec:det} and Section \ref{sec:sto}, deterministic source module and stochastic source module provide deterministic signal and stochastic signal as excitation signal respectively. To explore the influence of these two modules on predicted speech,  we visualize the deterministic signal, the stochastic signal and their spectrograms in Fig. \ref{fig:det_sto}.
It is easy to observe that the deterministic signal has obvious periodicity in the spectrogram, while it is difficult to find regularity in the spectrogram of stochastic signal, which is similar to the noise signal. These phenomena are consistent with the definition of deterministic signal and stochastic signal in Section \ref{sec:det} and Section \ref{sec:sto}, which proves that the structure and principle of our model are effective. The learned deterministic signal is not entirely a periodic glottal pulse with peak value. Because the neural network automatically learns the deterministic and stochastic signals, they can not be entirely consistent with the traditional source signal.
In addition, we find that the deterministic signal does not have a high correlation with the F0 of speech, which is different from the excitation signal  of traditional vocoder. This shows that in the neural vocoder, the periodic excitation signal does not have to be related to the $\text{F0}$, which can avoid the error accumulation caused by the  $\text{F0}$  extraction.

To further explore the role of the deterministic signal and stochastic signal, we conduct an ablation experiment.
 Two vocoders with the deterministic source module and the stochastic source module removed respectively are built, and their descriptions are as follows:
 \begin{itemize}
     \item \textbf{NDPS-MBE-woD} the NDPS-MBE vocoder removing the deterministic source module.
     \item \textbf{NDPS-MBE-woS} the NDPS-MBE vocoder removing the stochastic source module.
 \end{itemize}

 \begin{figure}[tp]
     \centering 
     \includegraphics[width=8cm]{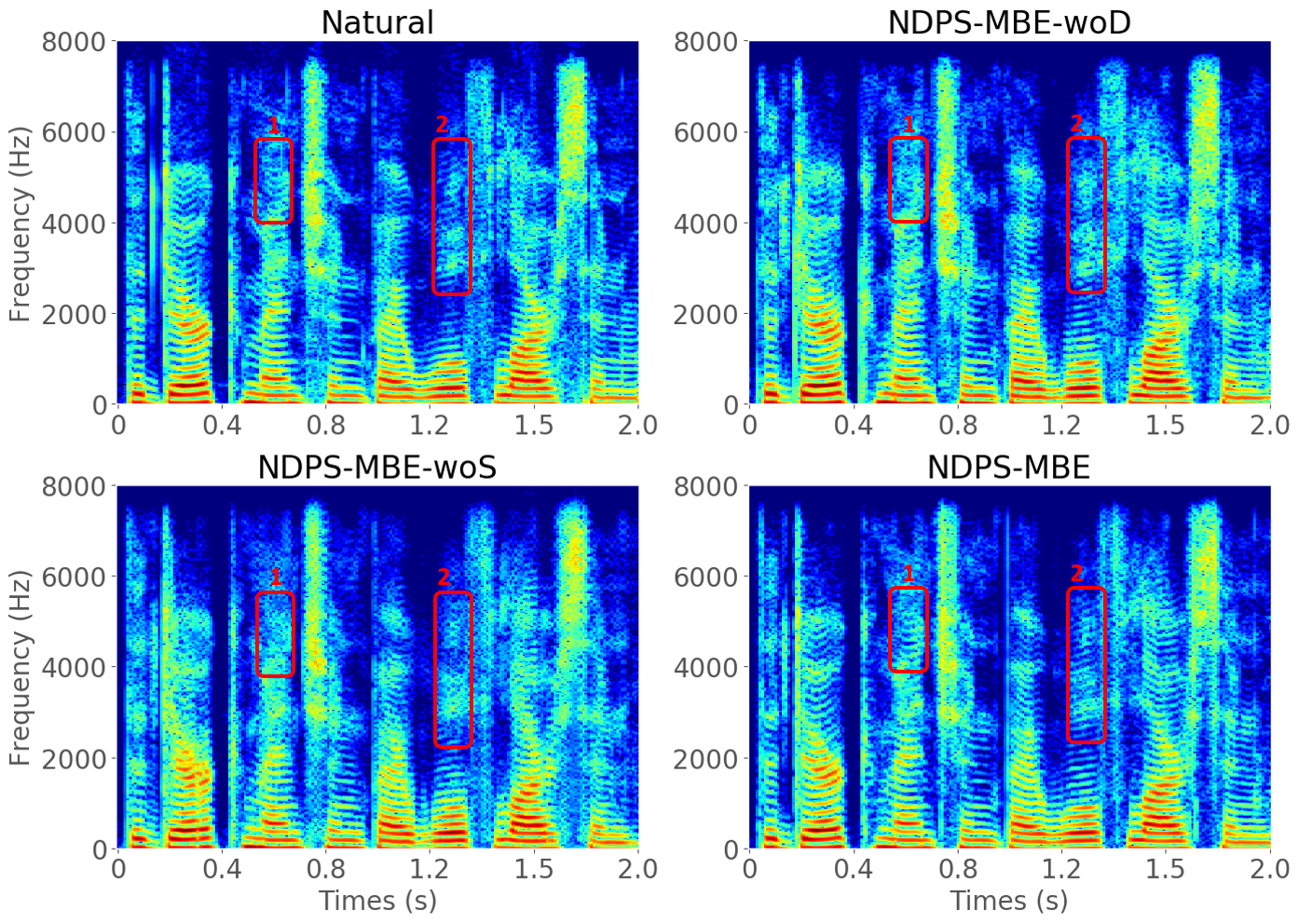}
     \caption{The spectrograms of the speech generated by different vocoders. 
Among them, the synthesized speech of model NDPS-MBE-woS has harmonic components, while the synthesized speech of model NDPS-MBE-woD is fuzzy in the high frequency part than that of model NDPS-MBE.}
     \label{fig:w_o_ds}
\vspace{-0.4cm}
 \end{figure}

Firstly, we compare the differences of synthesized speech spectrograms from different vocoders, which is shown in Fig. \ref{fig:w_o_ds}. It can be observed that when the stochastic signal is removed, there will be a lot of harmonics in the synthesized speech. These phenomena will lead to the decline of speech quality.  While the deterministic component is removed, the neural filter module completes the whole speech modeling process, that is, modeling the source and filter module simultaneously, which makes modeling difficult. For example, we can find NDPS-MBE is clearer by observing parts 1 and 2 of Fig. \ref{fig:w_o_ds}, while there is some slight blur in NDPS-MBE-woD. These can show that the deterministic module can help to learn a more detailed, clear spectrum. Especially in the high-frequency part, since the deterministic component contains high-frequency periodic signals. On this basis, it is easy to learn the detailed characteristics of high-frequency signals.

Besides, we conduct an ABX on the three vocoders. The result is shown in Table \ref{table:wods}. It can be observed that when any component is removed, the quality of synthesized speech will decrease. Specifically, through the comparison of model NDPS-MBE-woD and model NDPS-MBE-woS, it can be found that the stochastic source module has a greater impact on the sound quality than the deterministic source module. This shows that the stochastic signal in the excitation signal is more important. Through the comparison of model NDPS-MBE-woD and model NDPS-MBE, it can be found that although there are more people who think NDPS-MBE is good than NDPS-MBE-woD, which proves that the deterministic  source module  can further improve the quality of synthesized speech. However, 40\% of listeners think they have no difference. We attribute this to that the details of high frequency may be not easy to perceive. The formant in the low-frequency region is obvious, and the model can learn this information even without a deterministic module. Most of the differences between models NDPS-MBE-woD and NDPS-MBE are reflected in the high-frequency region, which is less perceptible than low frequency region.

In summary, the deterministic  source module and stochastic  source module are indispensable to the model NDPS-MBE, and the stochastic  source module has a greater impact on the quality of synthesized speech.

\begin{figure}[t]
    \centering 
    \includegraphics[width=8cm]{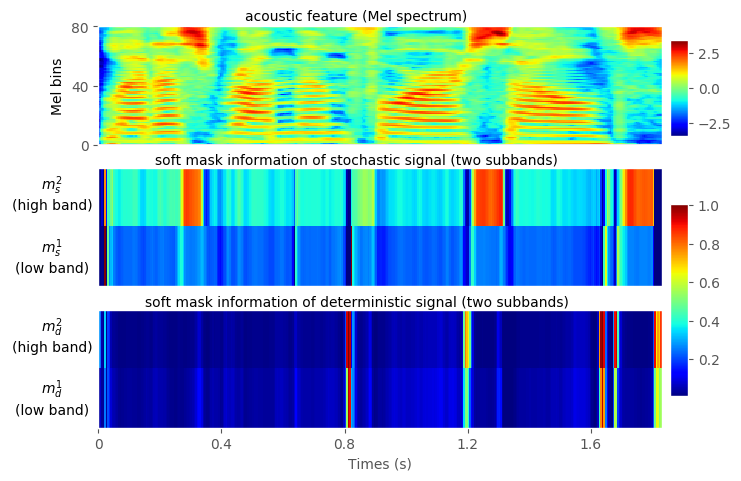}
    \caption{The input acoustic features (Mel spectrum) and the mask values predicted by the neural V/UV decision module. The mask values correspond to the mask information of the stochastic signal and the deterministic signal, and the excitation signal is decomposed into two subbands.
For example, $m_s^2$ represents the mask value corresponding to the second subband signal of the stochastic signal.}
    \label{fig:uvu_mask}
\vspace{-0.5cm}
\end{figure}

\subsubsection{The neural V/UV decision module}

As introduced in Section \ref{sec:uvu}, we add neural V/UV decision module to predict the soft mask value which can make the excitation signal more accurate and reduce the burden of neural filter. To explore whether neural V/UV decision module works, we first visualize the soft mask value predicted by the model, which is shown in Fig. \ref{fig:uvu_mask}.
For the deterministic signal, the difference between the mask values of the high-frequency part and the low-frequency part is small. Besides, the mask value of the voiced part is usually small, while  the mask value of the unvoiced part is larger. This shows that the network needs more deterministic components in the unvoiced part. For the stochastic signal, in the low-frequency part, we find that the mask value is more prominent in unvoiced, but smaller in voiced. While in the high-frequency part, the mask value is larger in voiced, but smaller in unvoiced. The two mask values are complementary to each other, which can model excitation signal well.

\begin{table}[t]
    \centering
        \caption{OBJECTIVE EVALUATION RESULTS OF  NDPS-MBE, NDPS-MBE-woVUV AND
NDPS-MBE-4 ON THE TEST SETS OF SPEAKERS $SLT$ WHEN USING NATURAL
ACOUSTIC FEATURES AS INPUT}
\scalebox{0.9}{
\begin{tabular}{c|ccc}
\hline \hline    & NDPS-MBE & NDPS-MBE-woVUV & NDPS-MBE-4 \\
\hline  SNR(dB)  & \textbf{4.7384} & 4.7060 & 4.4300\\
 LAS-RMSE(dB)    &   0.8005  &  0.8049 &  \textbf{0.7907}\\
  MCD(dB)    & \textbf{1.3073}  &  1.3710    &   1.3909\\ 
 F0-RMSE(cent)   & \textbf{9.3965} & 10.0466  &  10.2123\\
 V/UV error(\%)  & \textbf{2.8610}  &  3.4164  &  2.9896\\
\hline \hline
\end{tabular}}
\label{table:uvu}
\end{table}

Further, we remove the neural V/UV decision module from NDPS-MBE  to train a vocoder, which is denoted as \textbf{NDPS-MBE-woVUV}. The objective metrics on the $slt$ speaker dataset are shown in the Table \ref{table:uvu}.
By comparing model NDPS-MBE  and model NDPS-MBE-woVUV, we can find the influence of the neural V/UV decision module on the vocoder's performance. It can be found that after removing the   neural V/UV decision module, the objective metrics show a downward trend, especially the F0 and $\text{V/UV}$ metrics. This shows that the neural V/UV decision module has a positive role in predicting accurate F0 information.

\begin{table}[t]
\centering
\caption{AVERAGE PREFERENCE SCORES (\%) ON SPEECH QUALITY BETWEEN
NDPS-MBE  AND NDPS-MBE-woVUV MODEL WHEN USING
NATURAL ACOUSTIC FEATURES AS INPUT, WHERE N/P STANDS FOR “NO
PREFERENCE” 
}

\begin{tabular}{cccc}
\hline
\hline
NDPS-MBE & NDPS-MBE-woVUV &  N/P      & p \\ \hline
 \textbf{43.75 }               &        25.25                                    & 31.00   & \textless0.01  \\ \hline \hline
\end{tabular}
\label{table:vuvabx}
\vspace{-0.2cm}
\end{table}
We conducted a subjective ABX test to compare NDPS-MBE with NDPS-MBE-woVUV. The result is list in Table \ref{table:vuvabx}. It can be observed when the neural V/UV decision module is removed, the  speech quality has a decrease ($p \textless0.01$).

In summary, the neural V/UV decision module can learn effective soft mask value to help the model generate more accurate excitation signals. It can also improve the quality of synthesized speech.

\subsubsection{The multiband excitation strategy}
As introduced in Section \ref{sec:ndps-mbe}, we add the multiband excitation strategy to enrich the excitation signal. In Section \ref{sec:ndps-com},  we have compared the impact of multiband excitation  and found  it can  model the excitation signal better.
In this section, to explore the effect of the number of excitation bands on the vocoder's performance, we compare the following models.
\begin{itemize}
    \item \textbf{NDPS-MBE}: the excitation signal is decomposed into two subband signals ($M=2$).
    \item \textbf{NDPS-MBE-4}: the excitation signal is decomposed into four subband signals ($M=4$). Th configuration of the prototype filter $h_n(t)$ is shown in the Table \ref{table:1}.
\end{itemize}



\begin{table}[t]
\centering
\caption{AVERAGE PREFERENCE SCORES (\%) ON SPEECH QUALITY BETWEEN
NDPS-MBE  AND NDPS-MBE-4 MODEL WHEN USING
NATURAL ACOUSTIC FEATURES AS INPUT, WHERE N/P STANDS FOR “NO
PREFERENCE” 
}
\begin{tabular}{cccc}
\hline
\hline
NDPS-MBE & NDPS-MBE-4 & N/P & p \\ \hline
 \textbf{38.75}    & 24.75    & 36.50 &  \textless0.01   \\ \hline \hline
\end{tabular}
\label{table:mbeabx}
\vspace{-0.2cm}
\end{table}

The objective metrics and subjective evaluations are shown in Table \ref{table:uvu} and Table \ref{table:mbeabx} respectively. It can be found from Table \ref{table:uvu} that  the metrics of model  NDPS-MBE are generally better than NDPS-MBE-4, especially in the time domain metric SNR. By observing the impulse response of the analysis filter $h_k(n)$ in Fig. \ref{fig:1}, it can be found that when $M= 4$, the overlap between different frequency bands becomes larger, and it is difficult to eliminate these overlaps by setting parameters of Kaiser window $h(n)$.  We speculate that when the subband is decomposed into 4, there is too much overlap between different frequency bands, resulting in the excitation signal can not be well distinguished according to different frequency bands. From Table \ref{table:mbeabx}, it can be found that the performance of model NDPS-MBE is  better than that of model NDPS-MBE-4.

In summary, the number of subbands has a certain impact on the model's performance, and it is related to the parameter configuration of the analysis filter. Under the configuration of Table \ref{table:1}, the effect of decomposing the excitation signal into 2 subbands is better than decomposing into 4 subbands.
\vspace{-0.2cm}

\begin{figure}[tp]
    \centering 
    \includegraphics[width=7.5cm]{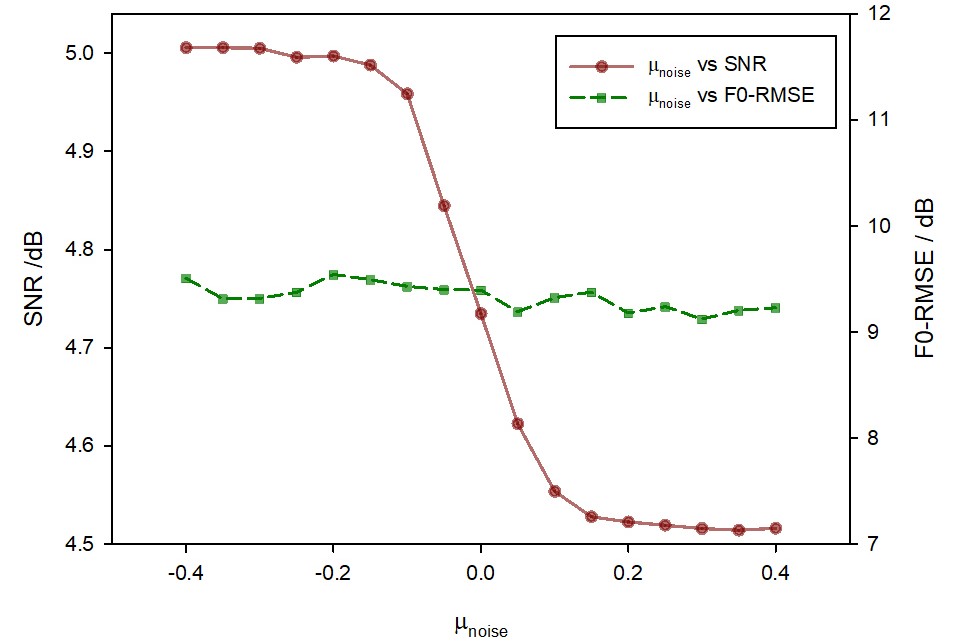}
    \caption{The $\text{SNR}$ and $\text{F0-RMSE}$ of synthesized speech with different $\mu_{noise}$.}
    \label{fig:unoise}
\vspace{-0.4cm}
\end{figure}
\subsection{Noise control}
As introduced in Section \ref{sec:noisecontrol}, we can modify the mask value of the stochastic signal to control the noise component in the synthesized speech. For the convenience of demonstration, we first modify the mask value of the full-band based on the model NDPS-MBE.
 We add a constant $\mu_{noise}$ to the mask value of all frequency bands of the stochastic signal.
To prevent the mask value from exceeding 1 or less than 0, which will cause no obvious changes 
to be observed, we set the range of $\mu_{noise}$ to $-0.4$ to $0.4$.
 The curves in Fig. \ref{fig:unoise} show the change of $\text{SNR}$ (red line) and $\text{F0-RMSE}$ (green line) of predicted speech with the change of $\mu_{noise}$. It can be observed that the $\text{F0-RMSE}$ has no obvious change trend with the increase of $\mu_{noise}$, but the $\text{SNR}$ has an obvious downward trend. This shows that the change of $\mu_{noise}$ can change the noise components in synthesized speech, which will not significantly affect the fundamental frequency signal, but can effectively control the signal-to-noise ratio of synthesized speech.
It is worth noting that the changing rate of SNR is greater in the interval $[-0.15,0.15]$.
 When $\mu_{noise}$ is greater than $0.15$ or less than $-0.15$, SNR changes slowly.  This is because some of the mask values may have reached saturation when the modification is large, and the change will be not obvious.

\begin{figure}[htp]
    \centering 
    \includegraphics[width=8.2cm]{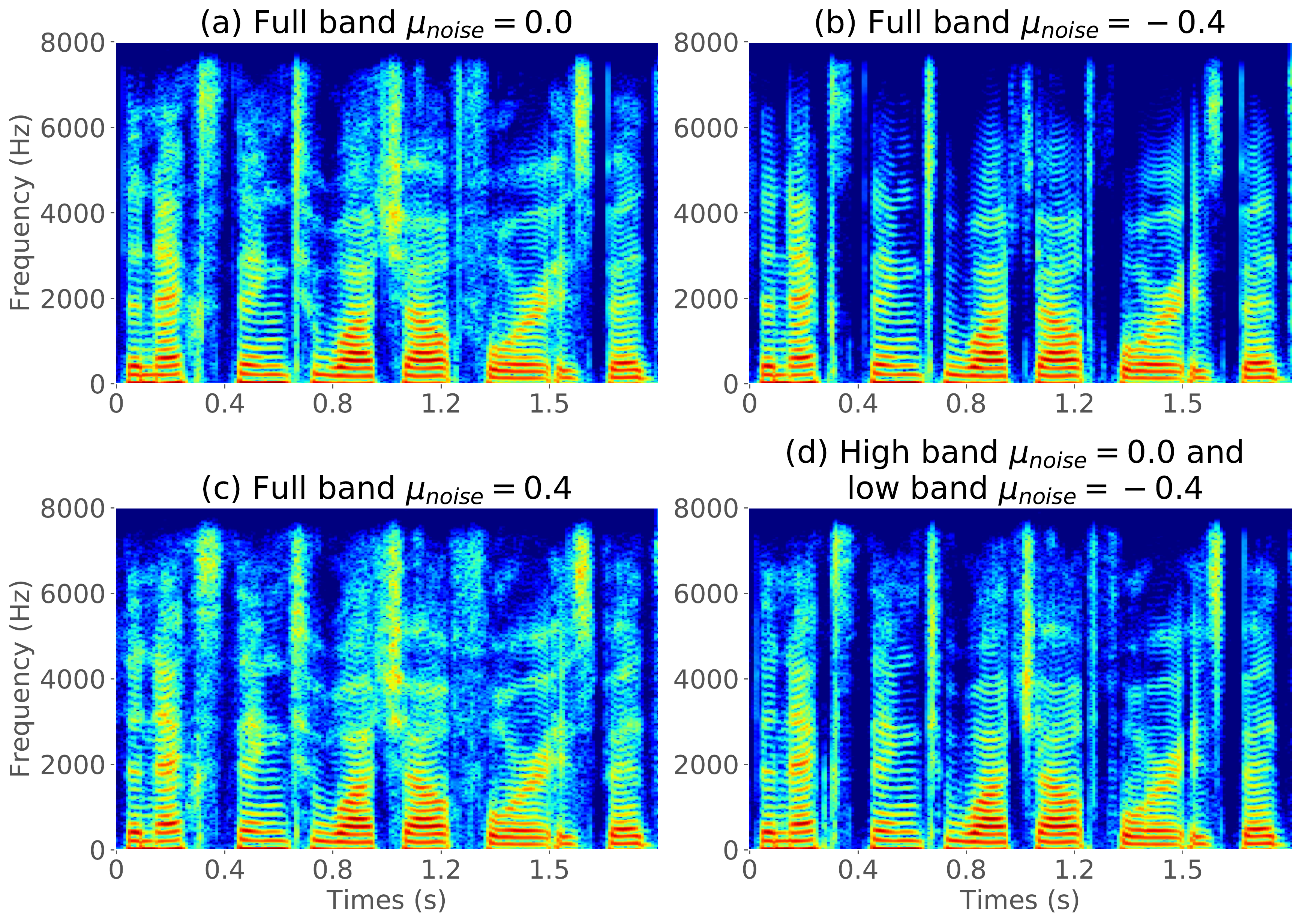}
    \caption{The spectrograms generated by  different noise control methods based on model NDPS-MBE.
Among them, the full band in (a), (b), and (c) means that the mask values of all frequency bands of the stochastic signal is added with $\mu_{noise}$. (d) means that the mask value of the low frequency band of the stochastic signal is modified, while the mask value of the high frequency band is unchanged.}
    \label{fig:uband}
\vspace{-0.1cm}
\end{figure}

In addition, to demonstrate the ability of noise control for the full frequency band or a specific frequency band,  Fig. \ref{fig:uband} shows the spectrum when different $\mu_{noise}$ values are applied to the full frequency band or a specific frequency band. By comparing the two subgraphs (a) and (b) in Fig. \ref{fig:uband}, when  $\mu_{noise}$ is reduced, the spectrum becomes clear and the noise component is less, and there is no unnatural problem through the audiometry. By comparing the three subgraphs (a), (b) and (c), we can find that when $\mu_{noise}$ becomes larger ($\mu_{noise}=0.4$), more stochastic components are introduced into the model, and more noise components are added to the synthesized speech. When $\mu_{noise}$ becomes smaller ($\mu_{noise}=-0.4$), less stochastic components are input into the model, and the synthesized spectrum is clearer. Subgraph (d) shows the ability to control only one special frequency band. We keep the mask value of high frequency band unchanged and reduce the mask value of low frequency band. It can be found that the spectrum of the low-frequency region becomes clear, but it also retains the noise component of the high-frequency part.

\vspace{-0.1cm}
\section{Conclusion}\label{sec:4}
This paper has proposed a novel neural vocoder named NeuralDPS, which has the characteristics of high speech quality, high synthesis efficiency and noise controllability. The NeuralDPS vocoder consists of a deterministic source module, a stochastic source module, a neural V/UV decision module, and a neural filter module. Besides, a multiband excitation strategy is employed to generate a more accurate excitation signal. A speech editing method is also proposed to adjust the predicted speech's SNR metric. As far as we know, this is the first neural vocoder that can modify the noise component of speech. The experimental results demonstrate that the NDPS-MBE model generates waveforms at least 280 times faster than the WaveNet-vocoder and it is 28\% faster than WaveGAN's synthesis efficiency on a single CPU core.  Besides, the quality of the synthesis speech is comparable to that from WaveNet.  In addition, we conducted detailed ablation experiments on the model to verify each module's role and improve the interpretability of the model. Finally, we show the effectiveness of editing speech based on the proposed framework. Decoupling the various components of the neural vocoder to achieve richer voice editing is the future work.

\vspace{-0.1cm}
\section{Acknowledgment}\label{sec:5}
This work is supported by the National Key Research and Development Plan of China (No.2020AAA0140002), the National Natural Science Foundation of China (NSFC) (No.62101553, No.61831022, No.61901473, No.61771472) and a Inria-CAS Joint Research Project (No. 173211KYSB 20190049).


%





\ifCLASSOPTIONcaptionsoff
  \newpage
\fi



%



\small
\bibliographystyle{IEEEtran}
\bibliography{refs}

%




%
\begin{IEEEbiography}[{\includegraphics[width=1.1in,height=1.25in,clip,keepaspectratio]{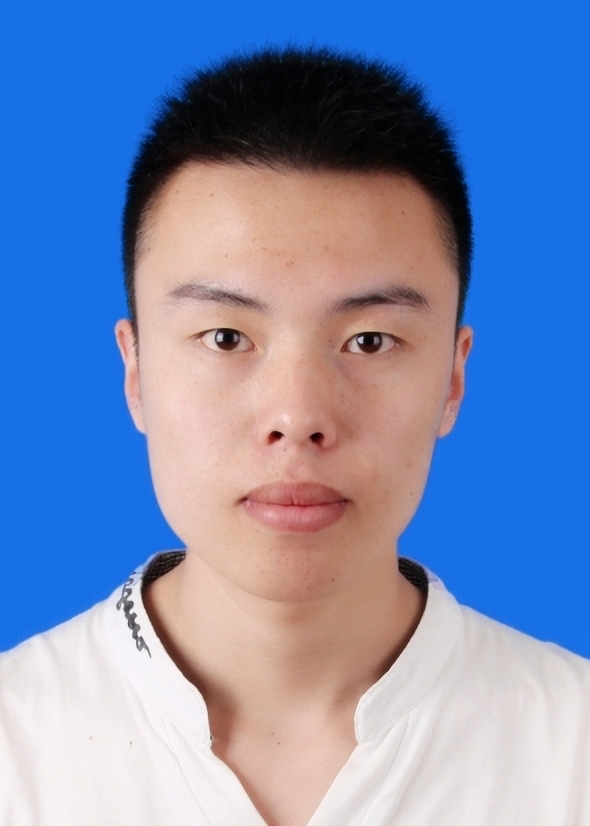}}]{Tao Wang}
received the B.E. degree from the
Department of  Control Science and Engineering, Shandong University (SDU), Jinan, China, in 2018. He is currently working
toward the Ph.D. degree with the National Laboratory
of Pattern Recognition, Institute of Automation (NLPR), Chinese
Academy of Sciences (CASIA), Beijing, China. His current
research interests include speech synthesis, voice conversion,  machine learning, and transfer learning.
\end{IEEEbiography}

\begin{IEEEbiography}[{\includegraphics[width=1.1in,height=1.25in,clip,keepaspectratio]{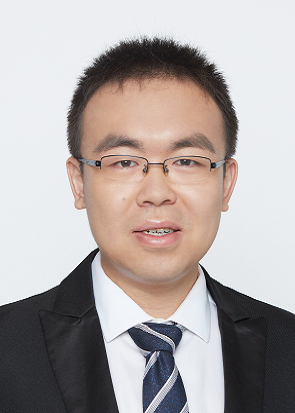}}]{Ruibo Fu}
is an assistant professor in National
Laboratory of Pattern Recognition, Institute of Automation, Chinese Academy
of Sciences, Beijing. He obtained B.E. from Beijing University of Aeronautics and Astronautics in 2015 and Ph.D. from Institute of Automation, Chinese Academy of Sciences in 2020. His research interest is speech synthesis and transfer learning. He has published more than 10 papers in international conferences and journals such as ICASSP and INTERSPEECH and has won the best paper award twice in NCMMSC 2017 and 2019. 
\end{IEEEbiography}

\begin{IEEEbiography}[{\includegraphics[width=1.1in,height=1.25in,clip,keepaspectratio]{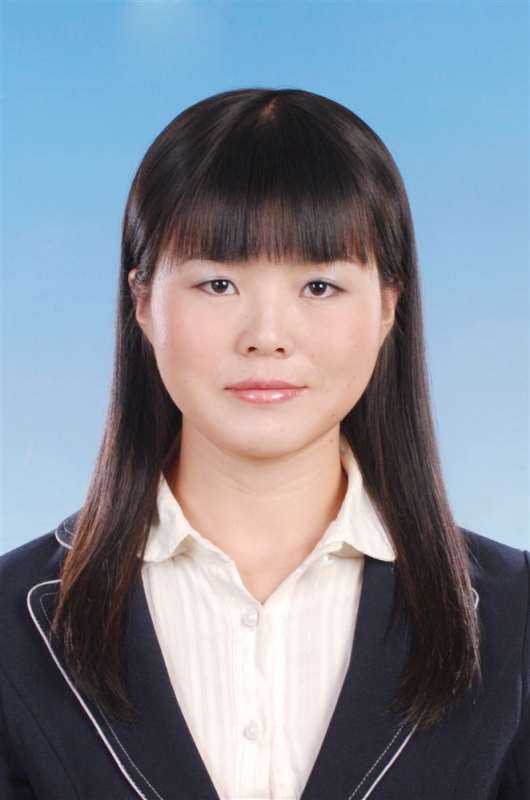}}]{Jiangyan Yi}
received the Ph.D. degree from the
University of Chinese Academy of Sciences, Beijing,
China, in 2018, and the M.A. degree from the Graduate
School of Chinese Academy of Social Sciences,
Beijing, China, in 2010. She was a Senior R\&D
Engineer with Alibaba Group from 2011 to 2014.
She is currently an Assistant Professor with the National
Laboratory of Pattern Recognition, Institute of
Automation, Chinese Academy of Sciences, Beijing,
China. Her current research interests include speech
processing, speech recognition, distributed computing,
deep learning, and transfer learning.
\end{IEEEbiography}

\begin{IEEEbiography}[{\includegraphics[width=1.1in,height=1.25in,clip,keepaspectratio]{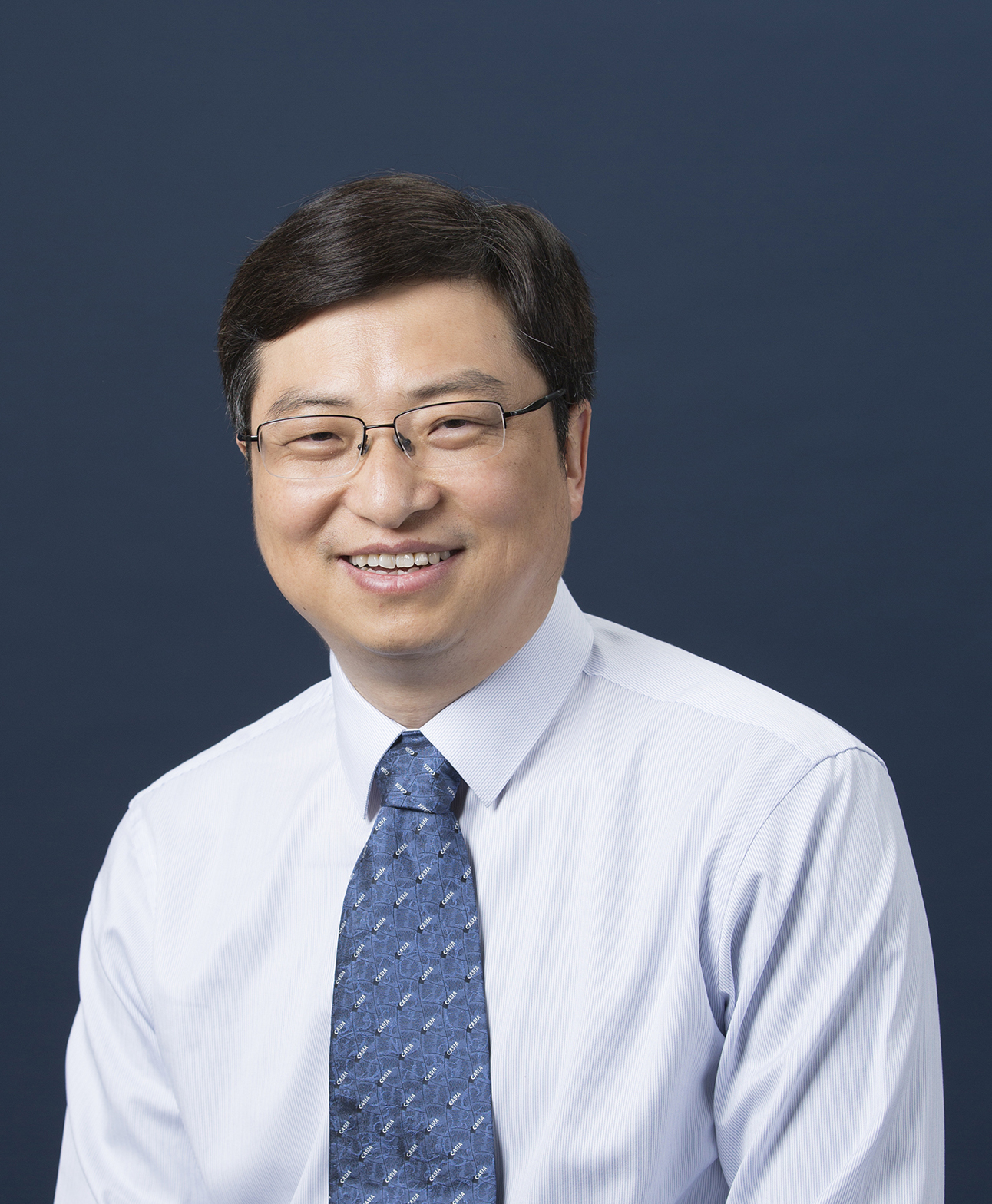}}]{Jianhua Tao}
(SM’10) received the Ph.D. degree from
Tsinghua University, Beijing, China, in 2001, and
the M.S. degree from Nanjing University, Nanjing,
China, in 1996. He is currently a Professor with
NLPR, Institute of Automation, Chinese Academy
of Sciences, Beijing, China. His current research interests
include speech synthesis and coding methods,
human computer interaction, multimedia information
processing, and pattern recognition. He has authored
or coauthored more than 80 papers on major journals
and proceedings including IEEE TRANSACTIONS ON
AUDIO, SPEECH, AND LANGUAGE PROCESSING, and received several awards
from the important conferences, such as Eurospeech, NCMMSC, etc. He serves
as the chair or program committee member for several major conferences,
including ICPR, ACII, ICMI, ISCSLP, NCMMSC, etc. He also serves as the
steering committee member for IEEE Transactions on Affective Computing, an
Associate Editor for Journal on Multimodal User Interface and International
Journal on Synthetic Emotions, the Deputy Editor-in-Chief for Chinese Journal
of Phonetics.
\end{IEEEbiography}

\begin{IEEEbiography}[{\includegraphics[width=1.1in,height=1.25in,clip,keepaspectratio]{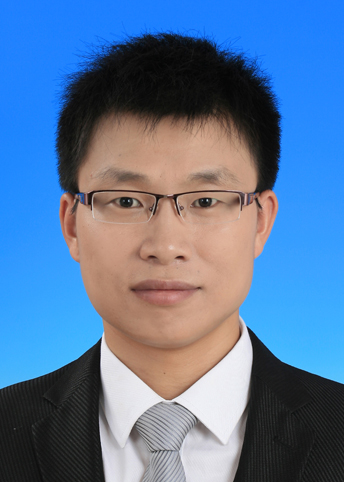}}]{Zhengqi Wen}
received the B.E. degree from the
Department of Automation, University of Science and
Technology of China, Hefei, China, in 2008 and the
Ph.D. degree from the National Laboratory of Pattern
Recognition, Institute of Automation, Chinese
Academy of Sciences, Beijing, China, in 2013. From
March 2009 to June 2009, he was an intern student
with Nokia Research Center, China. From December
2011 to March 2012, he was an intern student
with the Faculty of Systems Engineering,Wakayama
University, Japan. From July 2014 to January 2015,
he was a visiting scholar, under the supervision of Professor Chin-Hui Lee,
with the School of Electrical and Computer Engineering, Georgia Institute of
Technology, USA. He is currently an Associate Professor with the National
Laboratory of Pattern Recognition, Institute of Automation, Chinese Academy
of Sciences, Beijing, China.
\end{IEEEbiography}




\end{document}